\documentclass[a4paper]{aa}
\usepackage{graphicx,txfonts,aalongtable}

\def\teff{$T_{\rm eff}$}
\def\logg{$\log g$}

\def\vs{$v_{\rm e}\sin i$}
\def\ge{$g_{\rm eff}$}
\def\i{\,{\sc i}} \def\ii{\,{\sc ii}} \def\iii{\,{\sc iii}}
\def\hfs{{\it hfs}}
\newcommand{\vald}{{\sc vald}}
\newcommand{\dream}{{\sc dream}}
\newcommand{\bs}{$\langle B_{\rm s}\rangle$}
\newcommand{\kms}{km\,s$^{-1}$}

\newcommand{\figps}[1]{\resizebox{\hsize}{!}{\rotatebox{0}{\includegraphics{#1}}}}

\newcommand{\firps}[1]{\resizebox{\hsize}{!}{\rotatebox{-90}{\includegraphics{#1}}}}

\newcommand{\cm}{\,cm$^{-1}$}

\begin{document}

\title{Rare-earth elements in the atmosphere of the magnetic chemically peculiar star HD~144897%
\thanks{Based on observations collected at the European Southern Observatory,
Paranal, Chile (ESO programme No. 68.D-0254)}}%
\subtitle{New classification of the Nd\iii\ spectrum}

\author{T.~Ryabchikova\inst{1,2}
\and
A.~Ryabtsev\inst{3}
\and
O.~Kochukhov\inst{4}
\and 
S.~Bagnulo\inst{5}
}

\titlerunning{Rare-earth elements in the atmosphere of the magnetic chemically peculiar star HD~144897}
\authorrunning{T. Ryabchikova et al.}

\offprints{T. Ryabchikova, \\ \email{ryabchik@inasan.ru}}

\institute{Institute for Astronomy, University of Vienna,
T\"urkenschanzstrasse 17, A-1180 Wien, Austria 
\and
Institute of Astronomy, Russian Academy of Sciences, Pyatnitskaya 48, 109017 Moscow, Russia 
\and
Institute of Spectroscopy, Russian Academy of Sciences, 142190, Troitsk, Moscow region, Russia
\and
Department of Astronomy and Space Physics, Uppsala University Box 515, SE-751 20 Uppsala, Sweden
\and
European Southern Observatory, Casilla 19001, Santiago 19, Chile}

\date{Received / Accepted }

\abstract
{
The chemically peculiar stars of the upper main sequence represent a natural laboratory for the
study of rare-earth elements (REE).
}
{
We want to check the reliability of the energy levels and atomic line parameters for the second REE ions
currently available in the literature, and obtained by means of experiments and theoretical calculations.
}
{
We have obtained a UVES spectrum of a slowly rotating strongly magnetic Ap star, HD~144897, that
exhibits very large overabundances of rare-earth elements. Here we present
a detailed spectral analysis of this object, also taking into account effects of non-uniform vertical
distribution (stratification) of chemical elements.
}
{
We have determined the photospheric abundances of 40 ions. For seven elements (Mg, Si, Ca, Ti, Cr, Mn, Fe),
we have obtained a stratification model that allow us to produce a satisfactory fit to the observed profiles of
spectral lines of various strength. All the stratified elements, but Cr, show a steep decrease of concentration
toward the upper atmospheric layers; for Cr the transition from high to low concentration regions appears
smoother than for the other elements. REEs abundances, that for the first time in the literature have been determined
from the lines of the first \textit{and} second ions, have been found typically four dex larger than solar abundances.
Our analysis of REE spectral lines provide a strong support to the laboratory
line classification and determination of the atomic parameters. The only remarkable exception is Nd\iii,
for which spectral synthesis was found to be inconsistent with the observations. We have therefore 
performed a revision of the Nd\iii\ classification. We have confirmed the energies for 11 out of 24 odd energy levels
classified previously, and we have derived the energies for additional 24 levels of Nd\iii, thereby increasing
substantially the number of classified Nd\iii\ lines with corrected wavelengths and atomic parameters.
}
{}

\keywords{stars: magnetic fields -- stars: abundances -- stars: chemically peculiar -- stars: individual: HD 144897} 
	
\maketitle

\section{Introduction}
\label{intro}
 
Large overabundances of the rare-earth elements (REE) are the most typical characteristic of the
Upper Main sequence magnetic chemically peculiar (Ap) stars. In the past, REE studies were based on
the lines of the first ions, which are rather weak even in the spectra of REE-rich stars with
\teff$>$10000 K, and this makes blending a severe problem. In normal stars with these temperatures,
the lines of singly ionized REE are not visible at all. Therefore, the REE study was
limited by the first few most abundant elements, like La, Ce, Nd, Sm in some cases, and Eu, which
has few prominent lines in the optical region. The lines of the dominant second ionization stage
(REE3) were rarely studied quantitatively because of the lack of atomic data, although their
presence in the spectra of Ap stars was known for a long time (see, for example,  the pioneering
work by Swings (\cite{REE44}) on the REE3 line variations in $\alpha^2$\,CVn). A significant
progress in the laboratory works on the energy level classifications provides an opportunity for
theoretical calculations of the transition probabilities and other atomic parameters of spectral
lines. The most extensive calculations were done by the Belgium group at Mons University, who 
created a Database on Rare Earths at Mons University -- {\sc DREAM} (Bi\'emont et al.
\cite{dream99}). Other groups, which will be mentioned below, have also performed theoretical
calculations of the REE3 atomic structure, and this gives a possibility of comparing  different
results. At present, we have transition probabilities for all but unstable Pm\iii. Additionally,
thanks to great efforts of the Wisconsin University group (references will be given in the
corresponding Section), high precision laboratory measurements of the radiative lifetimes and
transition probabilities for the first REE ions became available.   These data give us a
possibility for the detailed study of the REE in Ap atmospheres.

However, the first applications of the REE3 atomic data have revealed some inconsistencies between
laboratory and calculated data  (wavelengths, oscillator strengths) and the observed lines in
stellar spectra. In particular, wavelengths of few Nd\iii\  lines calculated from the Nd\iii\
energy levels (Martin et al. \cite{NBS-REE78}) differ from the stellar ones by up to 0.1 \AA\ and
even more (see line identification list of HD~122970 spectrum in Ryabchikova et al.
\cite{RSHWH00}). This problem was resolved in a study of the laboratory Nd\iii\ spectrum by
Aldenius (\cite{Ald01}), who corrected several energy levels based on high precision wavelength
measurements. However, calculated oscillator strengths give discordant abundance results for some
Nd\iii\ lines in HgMn stars (Dolk et al. \cite{Dolk02}).

Ap stars provide a perfect natural laboratory for REE study. First, they have extreme
overabundances of these elements, therefore their spectra contain a large number of spectral lines
of both ionization stages. Second, magnetic splitting may give extra information to verify line
classification. And third, rapidly oscillating roAp stars have an outstanding pulsational
characteristic: in most stars only REE lines show large radial velocity pulsation amplitudes
(Savanov et al. \cite{SMR99}; Kochukhov \& Ryabchikova \cite{KR01a}, \cite{KR01b}), therefore we
can double-check whether a certain spectral feature belongs to REEs by looking at its pulsational
characteristics.

We chose HD~144897 as a template. This star has a surface magnetic field \bs\ varying between 8.5
and 9.5 kG over the period 48.43 d. (Mathys et al. \cite{bs97}). This field is strong enough to
produce clear Zeeman splitting of the spectral lines in unpolarized spectrum of  the slowly
rotating star,  and not too large to produce an overlapping of the Zeeman components from nearby
spectral lines. The quick look at the HD~144897 spectrum shows resolved Zeeman split features at
the positions of strong Pr\iii\ and Nd\iii\ lines. Furthermore, HD~144897 was never studied for
abundances, therefore our analysis will provide additional information for Ap-star chemistry. 

Our paper is structured as follows. Observations and data reduction are described in
Sect.\,\ref{observ}, the determination of the fundamental stellar parameters is presented in
Sect.\,\ref{parameters}.  The results of the magnetic spectrum synthesis in the homogeneous, as
well as in chemically stratified, atmosphere are given in Sect.\,\ref{abund}. Sect.\,\ref{nd3}
revises classification  of the Nd\iii\ spectrum. Discussion and conclusions are given in 
Sects.\,\ref{disc} and \ref{concl}, respectively.

\section{Observations and data reduction}
\label{observ}

High resolution, high signal to noise ratio UVES spectra of HD~144897
were obtained with the UVES instrument of the ESO VLT on 26 February 2002
within the context of program 68.D-0254. The UVES instrument is
described by Dekker et al.\ (2000). The observations were carried out
using both available dichroic modes. The detailed log of the
observations is given in Table~\ref{Table_UVES_Log}. In both the blue
arm and the red arm the slit width was set to 0.5$^{\prime\prime}$, for a spectral
resolution of about 80\,000.  The slit was oriented along the
parallactic angle, in order to minimize losses due to atmospheric
dispersion.  Almost the full wavelength interval from 3030 to 10400 \AA\
was observed except for a few gaps, the largest of which are at
5760-5835 \AA\ and 8550-8650 \AA.  In addition, there are several small
gaps, about 1\,nm each, due to the lack of overlapping between the
\'{e}chelle orders in the 860U setting.

The UVES data have been reduced with the automatic pipeline described in
Ballester et al.\ (2000).  For all settings, science frames are
bias-subtracted and divided by the extracted flat-field, except for
the 860 setting, where the 2D (pixel-to-pixel) flat-fielding is used,
in order to better correct for the fringing. Because of the high flux
of the spectra, we used the UVES pipeline \textit{average extraction}
method.

All spectra were normalized to the continuum with the help of an
interactive procedure, which employed either a low-degree polynomial
or a smoothing spline function.

\begin{table}
\caption{Log of UVES observations of HD~144897}
\label{Table_UVES_Log}
\begin{center}
\begin{tabular}{ccc}
\hline 
\hline
    Date   &    UT     & Setting (nm) \\
\hline
2002-02-26 & 08:01:25  & 346 \\
2002-02-26 & 08:10:26  & 346 \\
2002-02-26 & 08:01:24  & 580 \\
2002-02-26 & 08:10:32  & 580 \\
2002-02-26 & 08:25:07  & 437 \\
2002-02-26 & 08:25:07  & 860 \\
\hline
\end{tabular}
\end{center}
\end{table}

\section{Fundamental parameters}
\label{parameters}

Initial estimate of the effective temperature and surface gravity of 
HD\,144897 was derived using photometric data in the Geneva system.
The measurements were taken from the online catalogue maintained
at Geneva Observatory (Burki et al. \cite{B05})\footnote{{\tt http://obswww.unige.ch/gcpd/ph13.html}}.
Hipparcos parallax, $\pi=4.77\pm1.28$~mas, (Perryman et al. \cite{PLK97}) and
galactic coordinates of the star suggest that its reddening is significant.
From a variety of sources (Lucke \cite{L78}; Hakkila et al. \cite{HMS97};
Schlegel et al. \cite{SFD98}) we obtained $E(B-V)$ in the range from 0.13 to 0.48.
Large reddening is confirmed by the presence of very strong NaD interstellar absorption lines.
In this situation, the best approach is to make use of the reddening-free $X$, $Y$
parameters of the Geneva photometry. Using the calibration of K\"unzli et al. (\cite{KNK97}),
we determined $T_{\rm eff}=12051$~K and $\log g=4.34$. Subsequent correction for
the anomalous flux distribution of Ap stars (Hauck \& K\"unzli \cite{HK96})
yields $T_{\rm eff}=11100$~K. Stellar parameters were further adjusted to
fit the hydrogen H$\alpha$ and H$\beta$ lines. Finally, we have obtained
$T_{\rm eff}=11250$~K and $\log g=4.0$. Model atmosphere of HD\,144897
was calculated with the ATLAS9 code (Kurucz \cite{atlas9}), using ODF
with 10 times solar metallicity and microturbulent velocity 4~km\,s$^{-1}$
to mimic the enhancement of line opacity due to magnetic intensification.

We estimated the averaged surface magnetic field \bs\ by measuring distance between the resolved
Zeeman components of either pure doublets (Fe\ii\ 3303.46, 6149.48~\AA, Cr\ii\ 3421.20~\AA) or
pseudo-triplets (Fe\i\ 4260.47~\AA, Nd\iii\ 6145.07~\AA, Cr\ii\ 6231.68, 6275.87~\AA). Zeeman
components were approximated by the sum of Gaussians. The results for both groups are similar, and the
final value is \bs\,=\,8.79$\pm$0.10~kG. Our observations (HJD=2452331.83) correspond to a rotation
phase 0.13 for the ephemeris given in  Mathys et al.~(\cite{bs97}), and the derived \bs\ is fully
consistent with their magnetic curve for HD~144897. 

The projected rotational velocity was estimated using rather weak magnetically insensitive line
Fe\ii\ 6586.70~\AA\ (mean Land\'e factor \ge=0.02) and medium intensity Fe\ii\ 4491.40~\AA\
(\ge=0.4). Both lines indicate \vs\,$\approx$\,4~\kms\, which seems a little too large to
reproduce clearly visible Zeeman structure of many weak lines. A better fit of the synthetic
spectrum to the observations is provided by \vs\,=\,3~\kms, therefore we adopted this value in all
abundance analysis calculations.  

\section{Abundance analysis} 
\label{abund} 

A very important part of all abundance calculations is accurate atomic parameters of spectral
lines. Most atomic data in our study are taken from the \vald\ database (Kupka et al.
\cite{vald299}), which has been updated recently to include the new line lists: Ti\ii\ (Pickering
et al. \cite{Ti2-01}), Cr\ii, Fe\ii, Co\ii\ (Raassen \& Uylings \cite{RU98}), Mn\ii\ (Kling \&
Griesmann \cite{KG00}; Kling et al. \cite{KSG01}). The \dream\ REE line database (Bi\'emont et
al. \cite{dream99}) is also made accessible via the \vald\ extraction procedures. For the second
REE ions Land\'e factors were calculated by Quinet et al. (\cite{REE3-Z-04}). For Sr, Y, Zr, Ba,
Re Land\'e factors were taken from the ``Atomic Energy Levels'' tables by Moore (\cite{Moore71}).
For light elements O, Mg, Si Land\'e factors were calculated in the LS approximation. For Si\ii\
new experimental data on the Stark broadening and transition probabilities are available (Wilke
\cite{Wilke03}), which allows us to increase the number of Si\ii\ lines included  in the
abundance analysis. The new Si\ii\ oscillator strengths are smaller by 0.05 dex on average
compared to the values in \vald\ (see also Ryabchikova et al. \cite{RLK05}). This
difference does not influence the results of both standard and stratification analysis. Detailed
description of the REE atomic parameters will be given below.  

\subsection{Spectrum synthesis}
\label{synth}

Computations of the synthetic spectra of HD\,144897 were carried with the help of the magnetic
spectrum synthesis code \mbox{\sc synthmag\_fast}, which represents an improved version of the program
described by Piskunov (\cite{P99}). In this spectrum synthesis approach, the polarized radiative transfer
equation is solved numerically for  seven values of the angle between the line of sight and the normal to the
stellar surface. Resulting intensity spectra are integrated for a given rotational velocity. 
\mbox{\sc synthmag\_fast} uses a simplified model of the stellar magnetic field geometry with a
homogeneous surface distribution of both the field strength and field orientation. Although not
suitable for detailed analysis of polarization spectra, this model is adequate for the purpose of
fitting magnetically enhanced and splitted lines in the Stokes $I$ spectrum. In our study \mbox{\sc
synthmag\_fast} was used to calculate spectra both for homogeneous and stratified vertical
distribution of chemical abundances. The modelling of the lines with  different Zeeman patterns
require a mean inclination of the magnetic vector to the stellar surface $\sim$\,50--60\degr\ to
reproduce relative intensities of the $\pi$ and $\sigma$ components.  

The final abundances calculated for the magnetic field \bs=8.8 kG are presented in
Table\,\ref{Abund}.  Full list of the measured lines is available in electronic form only
(Table\,\ref{lines} of Online material). In Table~\ref{Abund} abundances derived for HD\,144897 are
compared with the recently published abundances of the two stars with similar effective temperatures:
HD~170973 (Kato \cite{Kato03}), and HD~116458  (Nishimura et al. \cite{HR5049}). The recent solar
photospheric abundances (Asplund et al. \cite{NSA05}) are given in the last column. With a few
exceptions abundances in all 3 Ap stars are  similar, in particular, for the REE.

\begin{table}[htb]
\caption{Element abundances in the Ap star HD\,144897 based on $n$ spectral lines. For elements with
the stratification analysis an abundance which provides the best fit to $n$ line profiles 
in homogeneous atmosphere is given. Abundances in the atmospheres of Ap stars HD\,170973, HD\,116458 
and in the solar photosphere (Asplund et al. \cite{NSA05}) are given for comparison. Results 
obtained from only one line are marked by ``:''. }
\label{Abund}
\begin{scriptsize}
\begin{center}
\begin{tabular}{llr|l|l|c}
\noalign{\smallskip}
\hline
\hline
Ion &\multicolumn{2}{c|}{HD\,144897}&HD\,170973&HD\,116458&Sun\\ 
    &$\log (N/N_{\rm tot})$ & $n$ &$\log (N/N_{\rm tot})$ & $\log (N/N_{\rm tot})$ &$\log (N/N_{\rm tot})$\\
\hline
\ion{~O}{i}  &-4.0     & 3  &-2.92   &-3.71  &~-3.38\\
Mg      &-5.6	& 3  &-4.87   &-4.80  &~-4.51\\
Si      &-3.8	&11  &-3.30   &-4.27  &~-4.53\\
Ca      &-6.0	& 5  &-5.36   &       &~-5.73\\
Ti      &-6.4	&16  &-4.56   &-6.53  &~-7.14\\
Cr      &-4.3	&23  &-4.48   &-4.54  &~-6.40\\
Mn      &-5.3	& 8  &-5.40   &-5.72  &~-6.65\\
Fe      &-3.5     &31  &-2.91   &-3.72  &~-4.59\\
\ion{Co}{i}  &-4.67(12)& 3  &        &-3.59  &~-7.12\\
\ion{Co}{ii} &-4.51(32)& 7  &        &-3.67  &~-7.12\\
\ion{Sr}{ii} &-7.45(55)& 4  &-6.18   &-9.06  &~-9.12\\
\ion{~Y}{ii} &-7.70(09)& 3  &	     &-7.69  &~-9.83\\
\ion{Zr}{ii} &-7.38(15)& 8  &-7.45   &	     &~-9.45\\
\ion{Ba}{ii} &-9.0:    & 1  &-9.42:  &-9.92: &~-9.87\\
\ion{La}{ii} &-7.43(12)& 4  &-7.92:  &-8.00: &-10.91\\
\ion{Ce}{ii} &-6.69(20)&10  &-6.87   &-7.34  &-10.46\\
\ion{Ce}{iii}&-6.64(18)& 8  &	     &       &-10.46\\
\ion{Pr}{ii} &-6.60(14)& 2  &-7.19   &-7.30  &-11.33\\
\ion{Pr}{iii}&-6.69(14)&14  &-6.87   &-7.32  &-11.33\\
\ion{Nd}{ii} &-6.45(12)& 9  &-6.48   &	     &-10.59\\
\ion{Nd}{iii}&-6.45(20)&13  &-6.63   &-7.26  &-10.59\\
\ion{Sm}{ii} &-6.98(21)& 4  &-7.07   &	     &-11.03\\
\ion{Sm}{iii}&-6.92(20)&10  &        &	     &-11.03\\
\ion{Eu}{ii} &-7.75(20)& 4  &-7.77   &-7.94  &-11.52\\
\ion{Eu}{iii}&-6.32(23)& 6  &        &-6.70  &-11.52\\
\ion{Gd}{ii} &-6.95(18)&13  &-7.17:  &-7.47  &-10.92\\
\ion{Gd}{iii}&-6.60(14)& 2  &	     & 	     &-10.92\\
\ion{Tb}{ii} &-7.83(10)& 3  &	     & 	     &-11.76\\
\ion{Tb}{iii}&-7.92(22)& 6  &-7.96   &	     &-11.76\\
\ion{Dy}{ii} &-7.12(22)&11  &-7.17   &-7.25: &-10.90\\
\ion{Dy}{iii}&-6.99(39)& 6  &-7.13   &-7.93: &-10.90\\
\ion{Ho}{ii} &-8.00(10)& 3  &	     &       &-11.53\\
\ion{Ho}{iii}&-8.08(16)& 6  &-7.92   &	     &-11.53\\
\ion{Er}{ii} &-7.55(14)& 2  &-7.54:  &	     &-11.11\\
\ion{Er}{iii}&-7.21(14)& 7  &-7.79   &	     &-11.11\\
\ion{Tm}{ii} &-8.12(20)& 3  &	     &       &-12.04\\
\ion{Tm}{iii}&-7.70(20)& 3  &	     &       &-12.04\\
\ion{Yb}{ii} &-7.50:   & 1  &	     &       &-10.96\\
\ion{Yb}{iii}&-7.50:   & 1  &	     &       &-10.96\\
\ion{Lu}{ii} &-8.60:   & 1  &	     &       &-11.98\\
\hline	          		      
  \teff      &11250\,K & & 10750\,K & 10300\,K &  5777\,K \\		      
  \logg      &3.70  & & 3.50  & 3.81  &  4.44 \\		    
  \bs\ (kG)  &8.8   & & 0.0   & 4.7   &       \\		       
  \vs\ (\kms)&3.0  & & 4.0   & 8.0   &\\  		
\hline	          		      
\end{tabular}
\end{center}
\end{scriptsize}
\end{table}

\begin{figure}[!t]
\firps{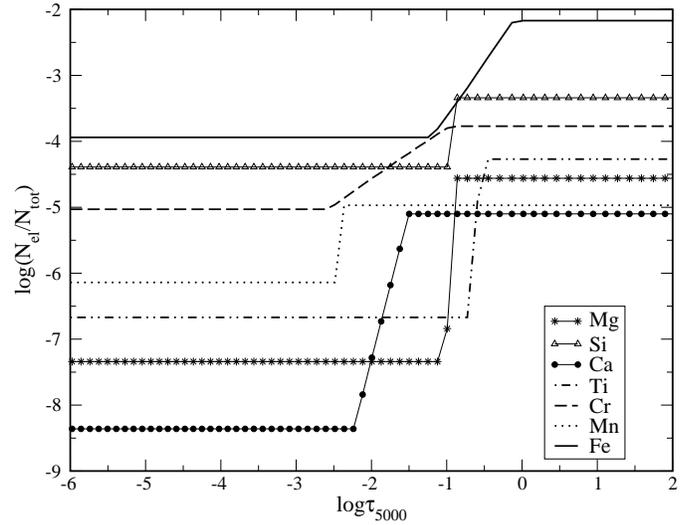}
\caption{Abundance stratification in the atmosphere of HD~144897.}
\label{distr}
\end{figure}

\subsection{Abundance stratification}
\label{strat}

\begin{figure*}[!th]
\centering
\includegraphics[height=75mm]{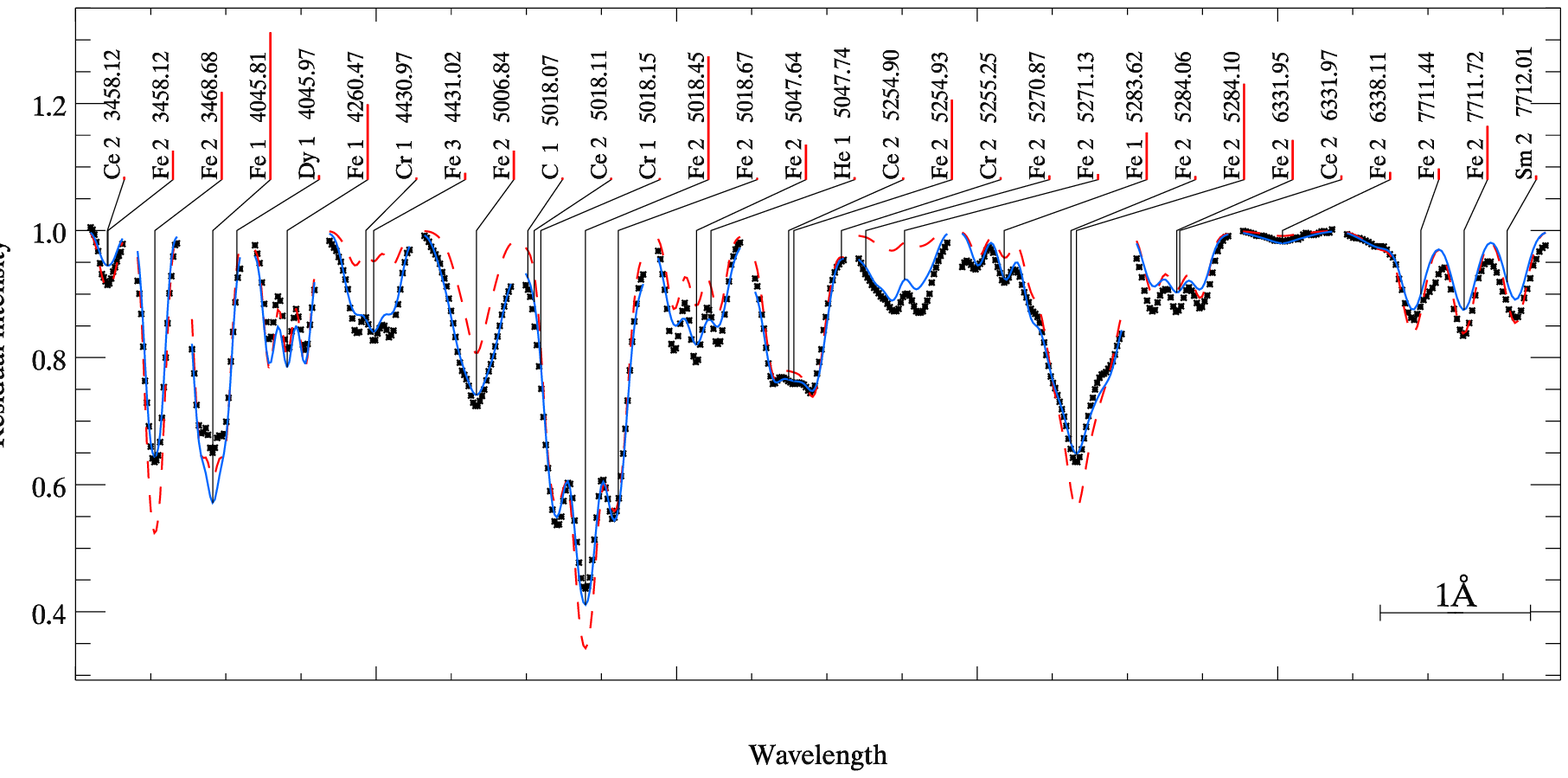}
%
\caption{A comparison between the observed  Fe line profiles and calculations with
stratified abundance distribution (full line)
and with homogeneous abundances (dashed line) from Table\,\ref{Abund}.  }
\label{Fe}
\end{figure*}

Preliminary analysis of the HD~144897 spectrum revealed  presence of Fe lines in three ionization
stages, with plenty of strong high-excitation Fe\ii\ lines. It was impossible to fit them with the
same abundance. Also, lines of Si\ii\ and Si\iii\ are present, and again it is not possible to
explain them with the same abundance. Therefore, we performed a stratification analysis of a few chosen
elements which have enough spectral lines of different strength, ionization (where possible),
excitation, sensitivity to Stark effect, and lying on both  sides of the Balmer Jump (BJ). For Mn
hyperfine splitting was taken into account using the data from Holt et al. (\cite{HSR99}).  The same step
function approximation of the abundance distribution which was adopted by Ryabchikova et al.
(\cite{RLK05}) in their analysis of HD~204411 was used here. The details of stratification analysis
are given in the cited paper. The only difference is that we used  magnetic spectrum synthesis for
HD~144897 instead of a regular spectrum synthesis for HD~204411. Stratification calculations were
performed for Mg, Si, Ca, Ti, Cr, Mn and Fe. First, we derived a mean abundance which gives the best
fit for all chosen line profiles, treated the same way as in stratification procedure, but with a
chemically homogeneous atmosphere. These homogeneous abundances are presented in Table\,\ref{Abund}
together with number of lines used in the analysis, and they were used as initial guess for the
stratification study. We also calculated mean deviation between  the observed and calculated line
profiles, which may be considered as a measure of the standard error. The obtained element
distributions are shown in Fig.\,\ref{distr}. 
An example of the fits to a sample of Fe line profiles representing a variety of Zeeman patterns is
given in Fig.\,\ref{Fe}. We do not expect to get a better fit of the Zeeman components due to a crude
magnetic geometry used in the magnetic spectrum synthesis. The improvement of the fit with the
stratified abundances over that with the homogeneous element distribution is also shown in
Fig.\,\ref{nd3-4914}.

Co\ii\ lines are very strong in HD~144897, especially in spectral region below BJ. They indicate an
overall Co abundance of $\log (Co/N_{\rm tot})=-4.2$:$-5.0$, i.e. more than 3~dex above the solar
photospheric abundance. However, the lack of hyperfine splitting (\hfs) data for Co\ii\ lines does not
allow us to  perform a detailed analysis for this element.

All stratified elements but Cr show a steep decrease of concentration towards the upper atmospheric
layers. The position of the abundance jump varies from $\log\tau_{5000}\approx-1$ (Mg, Si, Ti and Fe)
to $-2.5$ (Mn). The Mg distribution in HD~144897 differs from the distribution derived in HD~204411
(Ryabchikova et al. \cite{RLK05}) and in HD~133792 (Kochukhov et al. \cite{KTRM06}), where a slow
increase of the Mg abundance towards the upper atmosphere was obtained. We cannot consider our present
results for Mg as conclusive because of small number of Mg lines available for HD~144897. More stars
in a large temperature range should be investigated to get a better picture of the abundance
distribution variations with effective temperature and other stellar parameters. In HD~144897 the 
abundance jumps for Cr, Mn and Fe are not very large, $\sim$1--2~dex. The jump is larger for Fe than
for Cr and Mn.   As demonstrated by Shulyak et al. (\cite{LL04}), similar Fe distribution in the
hotter star CU~Vir  (12750 K) leads to a better modelling of the observed energy distribution, in
particular, the $\lambda$\,5200 \AA\ depression. Similar Ca and Cr distributions, but with slightly
smaller abundance jumps,  were derived in the atmospheres of cooler and more evolved stars  HD~204411
(8400 K) and HD~133792 (9400 K). Ca has the largest jump in all three stars. Si abundance jump is
smaller than in HD~204411 and HD~133792. It is roughly comparable with the Si distribution in
slightly hotter star HD~10221 (Glagolevskij et al. \cite{GRC05}). 

For all elements except Mg and Ca abundances are never below solar over the part of the atmosphere
where we can consider our results to be reliable. This atmospheric part is limited by the line
formation sensitivity to temperature and pressure. We cannot say much about abundances above 
$\log\tau_{5000}=-$4.

\subsection{Rare-earth elements}
\label{REE}

HD~144897 is the second object after the famous Przybylski's star (HD\,101065) for which abundances of
all REEs but Pm are derived, and the first star in which abundances for most REEs are derived from the
lines of the first and second ions. For odd isotopes hyperfine splitting is known and can be taken
into account. However, in the presence of magnetic field, the coupling between the Zeeman and \hfs\
components changes shape of spectral line and should be taken into account in spectral synthesis.
However, we noticed that in most cases the line profiles are fitted reasonably well with the pure
Zeeman splitting. The lines of the first ions of the REE are rather weak in  HD~144897 spectrum, and
the derived abundance should not be affected much by the \hfs\ effects. Consequently, we did not use
the \hfs\ data in our analysis, with the exception of Pr\ii, Eu\ii\ and Lu\ii. Below we discuss briefly each
element.

When no specific reference is given for  atomic data, it means that all relevant parameters 
are extracted from \vald\ (Kupka et al. \cite{vald299}).

\underline{\bf Lanthanum}. The oscillator strengths and \hfs\ constants for La\ii\ lines are taken
from Lawler et al. (\cite{La2}). Two lines of La\iii\ at $\lambda$~3171.69 and 3517.16 are seen in our
spectrum, but the lines are rather wide and are obviously affected by \hfs, which is unknown. 

\begin{figure}[th]
\figps{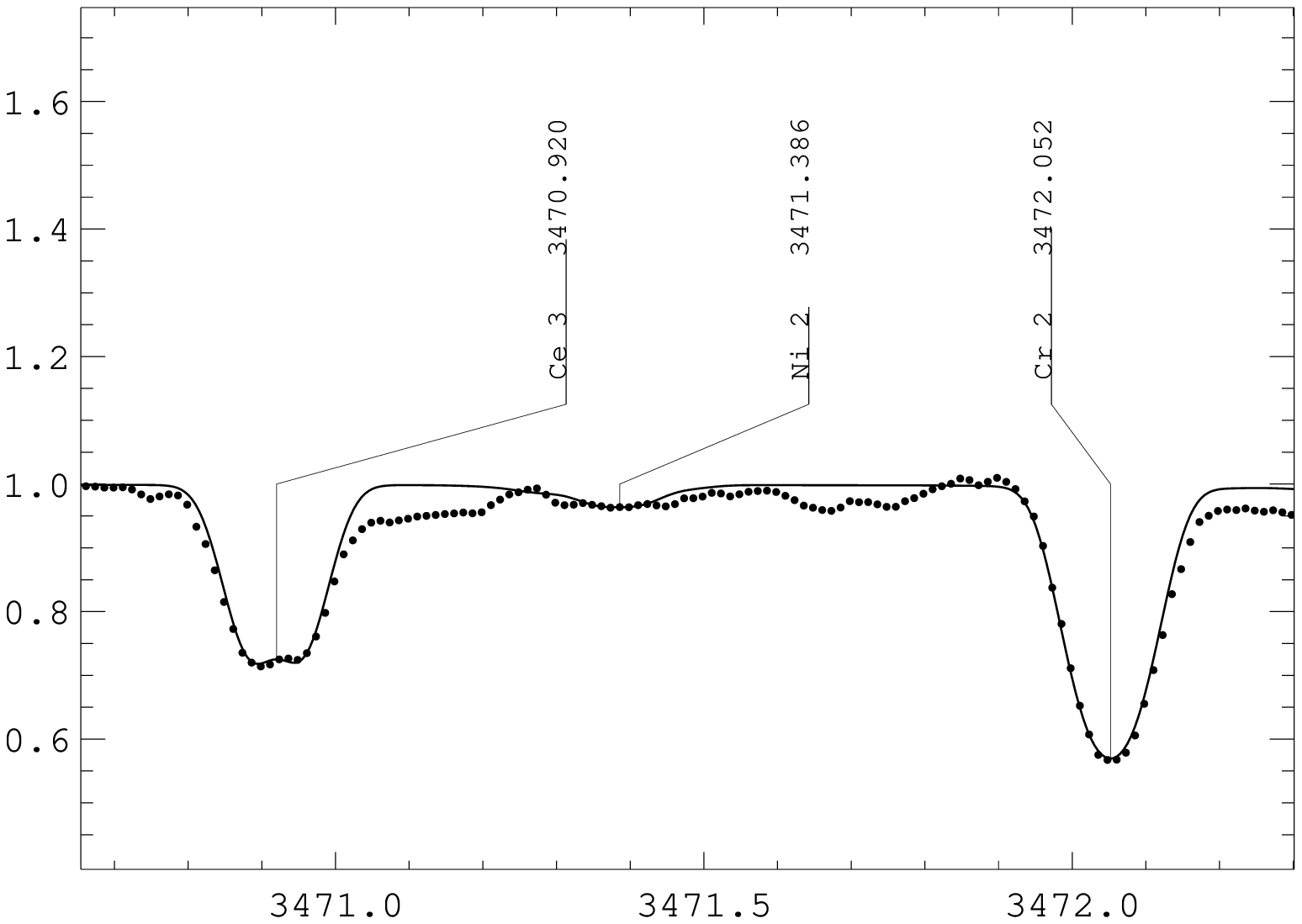}
\caption{The Ce\iii\ 3470.92~\AA\ line in the spectrum of HD~144897 (dots). 
Spectral synthesis is shown by full line.}
\label{ce3}
\end{figure}

\underline{\bf Cerium}. Oscillator strength for Ce\ii\ and Ce\iii\ lines are taken from Palmeri et al.
(\cite{Ce2}) and from Bi\'emont et al. (\cite{Ce3}), respectively. All lines are nicely reproduced
with adopted magnetic field. Figure\,\ref{ce3} shows a fit of the synthesized Ce\iii\ 3470.92~\AA\
line to the observed one. Most Ce lines in HD~144897 are well represented with the available atomic
data.     

\begin{figure}[th]
\figps{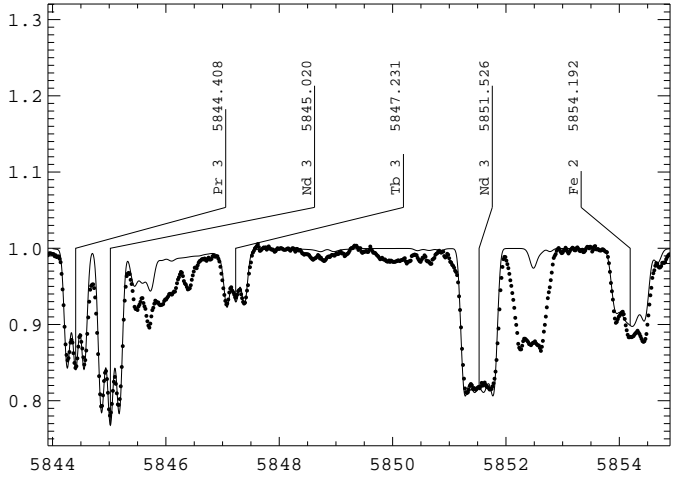}
\caption{Examples of the fit for Pr\iii, Nd\iii\ and Tb\iii\ lines (full line) to the observed HD~144897 spectrum (dots).}
\label{pr-nd-tb}
\end{figure}

\underline{\bf Praseodymium}. Atomic data from \vald\ and \hfs\ constants from Ginibre (\cite{Ginibre1989})
were used in the Pr\ii\ analysis, which was based on two  relatively weak lines at 
$\lambda\lambda$~4206.72, 5322.77~\AA. Other Pr\ii\ lines are in blends.
For Pr\iii\ lines new calculations by one
of us (AR) based on the extended energy levels given  by Wyart et al. (\cite{Pr3}) were applied. 
New calculations contain transition probabilities for the 4${\rm f}^3$\,$^4{\rm F}$ term, which
is missing in the \dream\ list (Bi\'emont et al. \cite{dream-Pr3}). However, a number of
intermediate strength lines from this term are observed in our spectrum. In general, a difference
between \dream's and present oscillator strengths for 10 lines in common does not exceed
0.08 dex. Calculated Land\'e factors represent nicely the observed Zeeman patterns (see
Fig.\,\ref{pr-nd-tb} for Pr\iii\ 5844.41~\AA).

\begin{figure}[th]
\figps{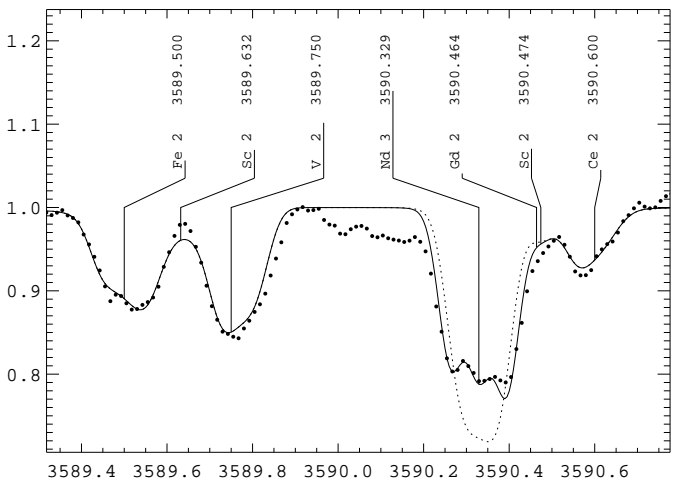}
%
\caption{The fit for Nd\iii\ lines  to the observed HD~144897 spectrum (dots) in the 
 3590~\AA\ region. Dashed line represents previous classification for the 
 Nd\iii\ 3590.24~\AA \ feature, while full line shows new classification for this line (see Sect.\ref{nd3}).} 
\label{nd3-3590}
\end{figure}

\begin{figure}[th]
\figps{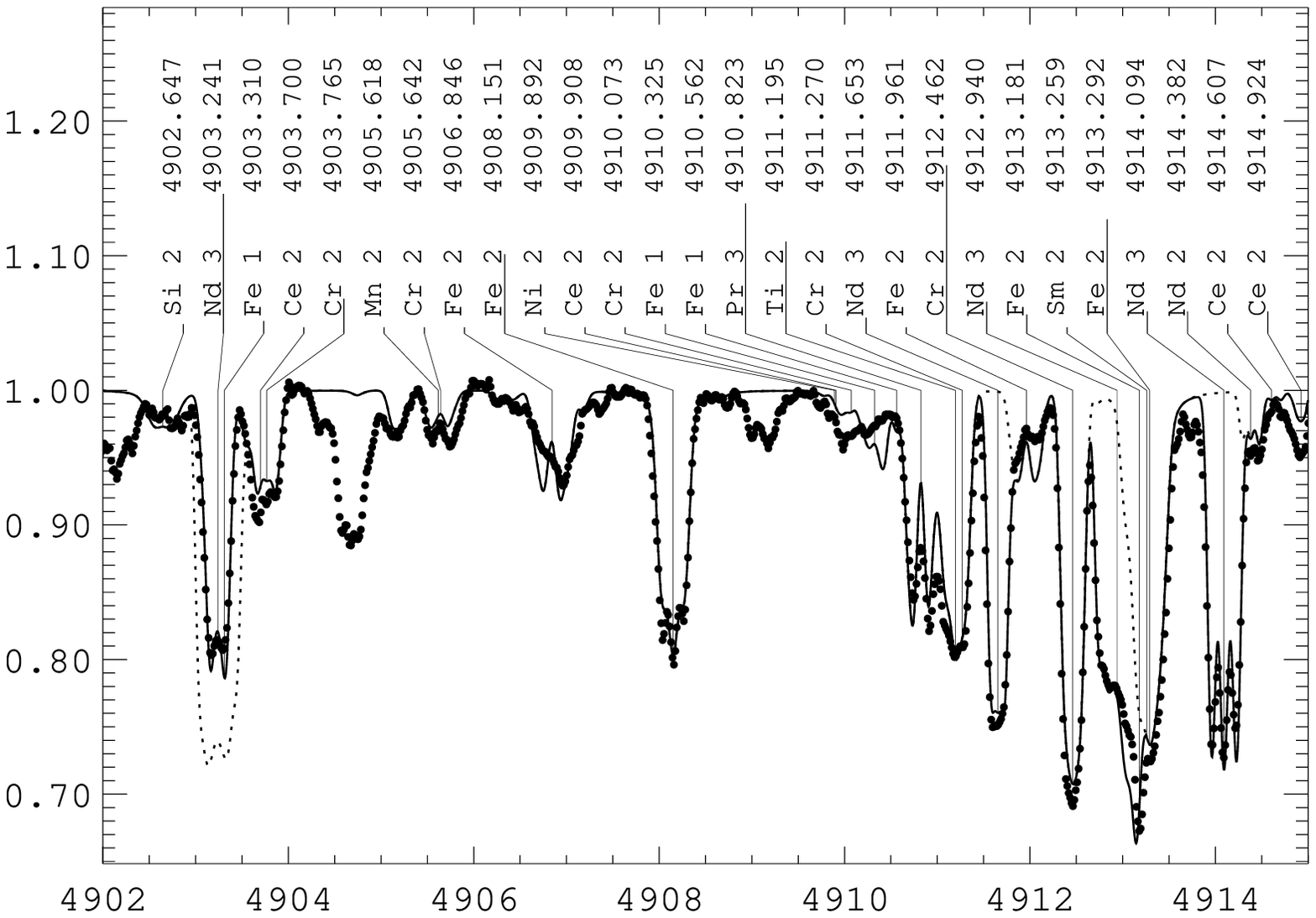}
\figps{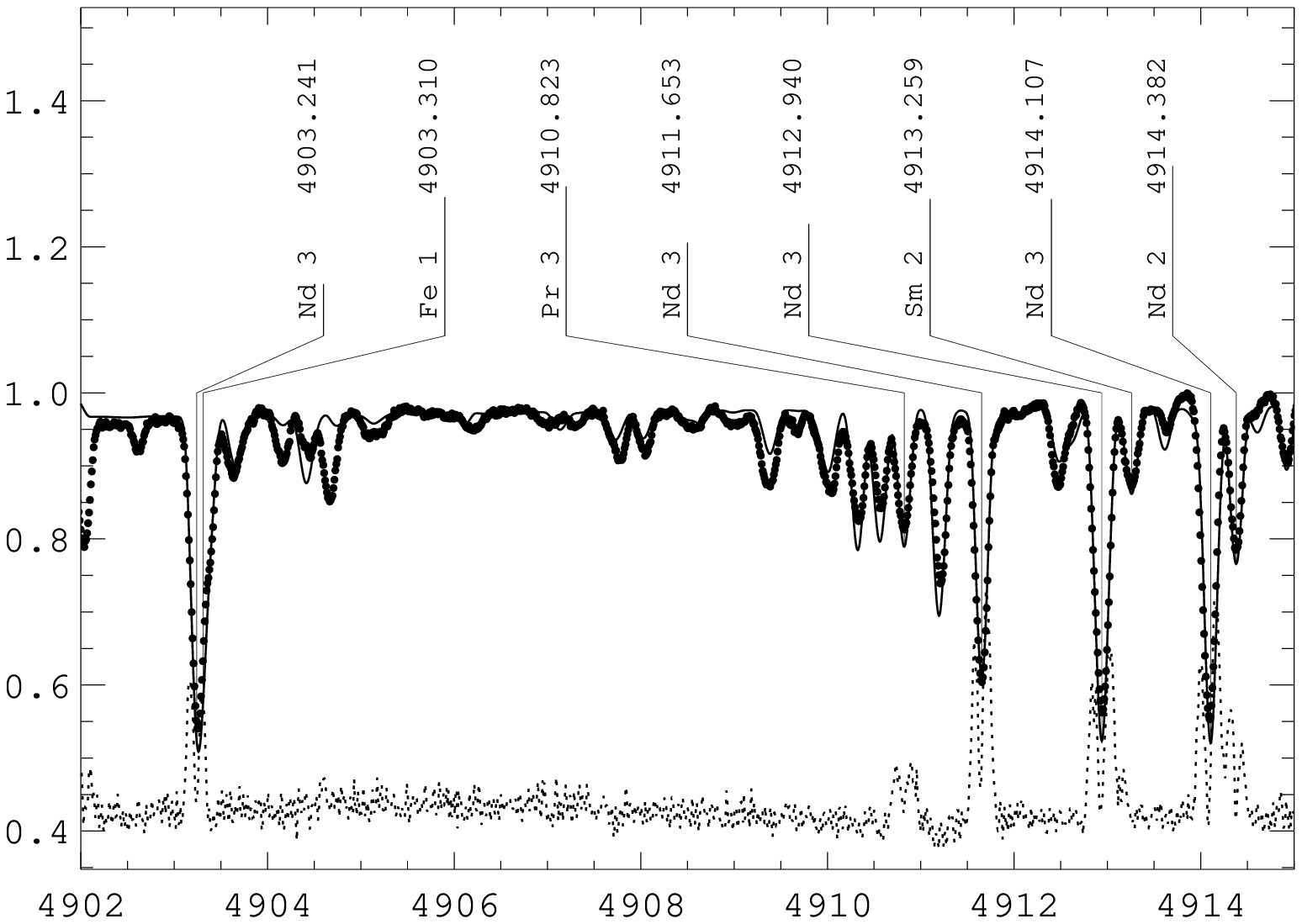}
\caption{The upper panel compares the observed spectrum of HD\,144897 (dots) in the
4902-4915~\AA\ region with 
the spectrum synthesis based on the old (dotted line) and revised (solid line) 
classification of the Nd\iii\ spectrum. The lower panel shows the same spectral interval
for the roAp star HD\,24712. The dotted curve gives standard deviation (arbitrary
scale) as a function of wavelength for the time-resolved observations of HD\,24712. 
Double-peaked bumps in the standard deviation indicate lines showing large-amplitude
radial velocity variation.} 
\label{nd3-4914}
\end{figure}

\underline{\bf Neodymium}. Recent experimental oscillator strengths (Den Hartog et al. \cite{Nd2-03})
were used to derive abundances from Nd\ii\ lines. Analysis of the Nd\iii\ lines met with difficulties,
because we did not find a number of spectral lines which should be visible according to \dream\ data.
For some spectral lines, for example, $\lambda\lambda$~3590.32, 4903.24~\AA\ the theoretical Zeeman pattern deviates
from the observed one (Figs.\,\ref{nd3-3590}, \ref{nd3-4914} ).
The new Nd\iii\ classification
will be considered in Sect.\,\ref{nd3}. We based our abundance determinations on the lines whose
Zeeman patterns are well reproduced by the Land\'e factors given in three independent sets of
theoretical calculations: Bord (\cite{Bord00}), Zhang et al. (\cite{Nd3}) and the present
study (Sect.\,\ref{nd3}). An example of the fit for the Nd\iii\ 5845.02~\AA\ line is shown in
Fig.\,\ref{pr-nd-tb}.    

\underline{\bf Samarium}. Accurate experimental data are available now for 958 Sm\ii\ lines (Lawler et
al. \cite{Sm2}), and we used them in our analysis. Note, that for the 4 lines studied here, the data
by Lawler at al. agree perfectly with the line parameters included in \vald. For Sm\iii\ lines
oscillator strengths are taken from Bi\'emont et al. (\cite{Sm3}). The authors give 3 sets of data:
experimental transition probabilities for a limited number of lines, relativistic Hartree-Fock (HF)
calculations combined  with the experimental branching ratios (BR), and HF calculations with the core
polarization effects included. The last data are available for the larger number of spectral lines,
and was used in our abundance analysis (Table\,\ref{Abund}).  With experimental data only we get $\log
({\rm Sm}/N_{\rm tot})=-7.03\pm0.23$ from 3 spectral lines. The HF calculations with the  experimental BR
give  $\log ({\rm Sm}/N_{\rm tot})=-6.90\pm0.14$ for 8 spectral lines. Abundance determinations with 3 sets
of oscillator strengths agree within the error limits. Calculated Land\'e factors for the lines used
in abundance analysis represent the observed Zeeman patterns well.      

\underline{\bf Europium}. Experimental oscillator strengths and \hfs\ constants for Eu\ii\ lines are
taken from Lawler et al. (\cite{Eu2}). 
For the Eu\iii\ lines we again used the
new transition probabilities and Land\'e factors calculations, which are based on the extended
analysis of the Eu\iii\ energy levels (Wyart et al. \cite{Pr3}). Generally, the new line parameters do
not differ significantly from the \dream\ data. Eu is the only rare-earth element  for which more than 1.0
dex difference between abundances from the lines of the first and second ions is obtained. The
NLTE effects may explain about half of the observed difference in the case of homogeneous Eu
distribution (see Mashonkina et al. \cite{NLTE-Eu}). NLTE effects of the same order are expected for
Nd. However, under the LTE assumption, Nd\ii\ lines should provide lower abundances to Nd\iii\ 
(L. Mashonkina, private communication), but we do not find this discrepancy for HD\,144897.    

\underline{\bf Gadolinium}. The Gd\iii\ oscillator strengths are taken from Bi\'emont et al.
(\cite{Gd3}). The two lines, $\lambda\lambda$~3118.04 and 3176.66~\AA, although partially blended, are
synthesized satisfactorily with the theoretical Land\'e factors. 

\underline{\bf Terbium}. Experimental data for Tb\ii\ lines are adopted from Lawler et al.
(\cite{Tb2}). For Tb\iii\ the transition probabilities and Land\'e factors were calculated by one of
us (AR) based on the extended analysis of the Tb\iii\ energy levels (Wyart et al. \cite{Pr3}).
Figure\,\ref{pr-nd-tb} gives an example of the fit to the observed Tb\iii\ 5847.23~\AA\ line. All
other unblended lines in our analysis are also fitted perfectly with the  current atomic data. The
Tb\iii\ transition probabilities were also calculated by Bi\'emont et al. (\cite{Tb3}), and the two
sets of parameters agree within 0.05~dex. 

\begin{figure}[!t]
\figps{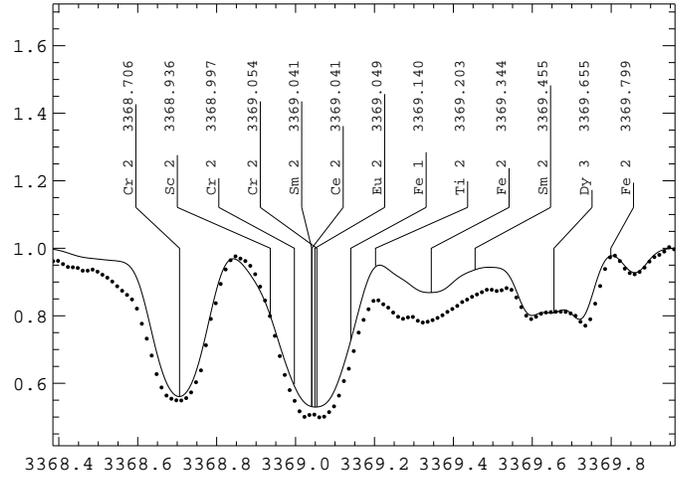}
\caption{The same as in Fig.\,\ref{pr-nd-tb} but for Dy\iii\ line.} 
\label{dy3}
\end{figure}

\underline{\bf Dysprosium}. Experimental data for Dy\ii\ lines (Wickliffe et al. \cite{WLN00}) and
theoretical calculations  for Dy\iii\ lines (Zhang et al. \cite{Dy3}) were used in the present study.
Figure\,\ref{dy3} shows the fit for one of the Dy\iii\ lines at $\lambda$~3669.66~\AA. Although a
standard deviation in abundance obtained from Dy\iii\ lines is larger than for other species, a good
agreement between abundances derived from both ionization stages provides a support for relatively
accurate theoretical atomic parameters of Dy\iii\ lines.

\begin{figure}[th]
\figps{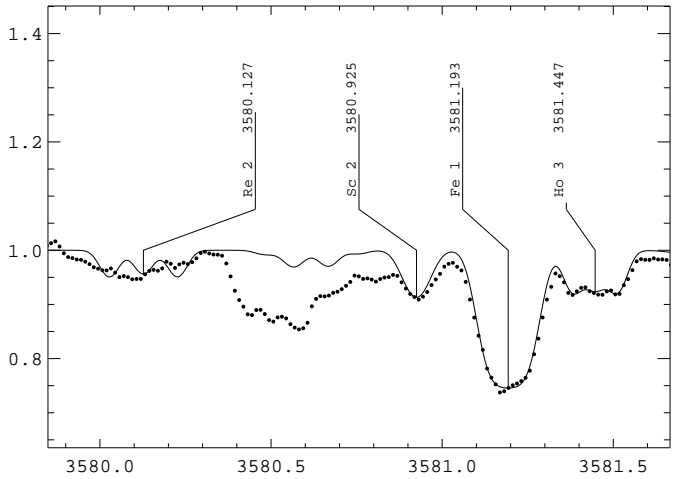}
\caption{The same as in Fig.\,\ref{pr-nd-tb} but for Ho\iii\ line.}
\label{ho3}
\end{figure}  

\underline{\bf Holmium}.  Experimental data for Ho\ii\ lines are taken from Lawler et al.
(\cite{Ho2}). Improved partition function for Ho\ii\ (Bord \& Cowley \cite{HoPF}) were implemented in
{\sc synthmag}. Ho\iii\ lines were synthesized using oscillator strengths calculated by Zhang et al.
(\cite{Ho3}) and normalized to the measured  lifetimes. Figure\,\ref{ho3} demonstrates the fit of the
calculated Ho\iii\ 3581.45~\AA\ line (the strongest in our study) to the observed spectrum. Obviously, 
\hfs\ effects are not significant. The fit of similar quality is obtained for other Ho\iii\
lines.   

\begin{figure}[th]
\figps{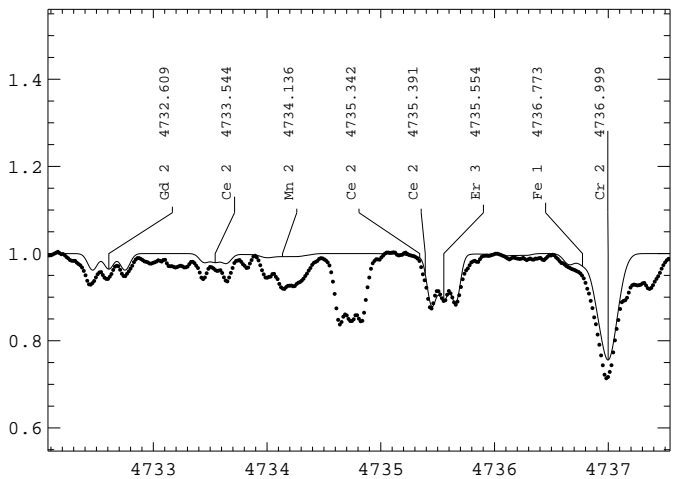}
\caption{The same as in Fig.\,\ref{pr-nd-tb} but for Er\iii\ line.}
\label{er3}
\end{figure}

\underline{\bf Erbium}.  For Er\iii\ calculations we used oscillator strength from Bi\'emont et al.
(\cite{Er3}). Figure\,\ref{er3} illustrates the quality of existing Er\iii\ data.

\begin{figure}[th]
\figps{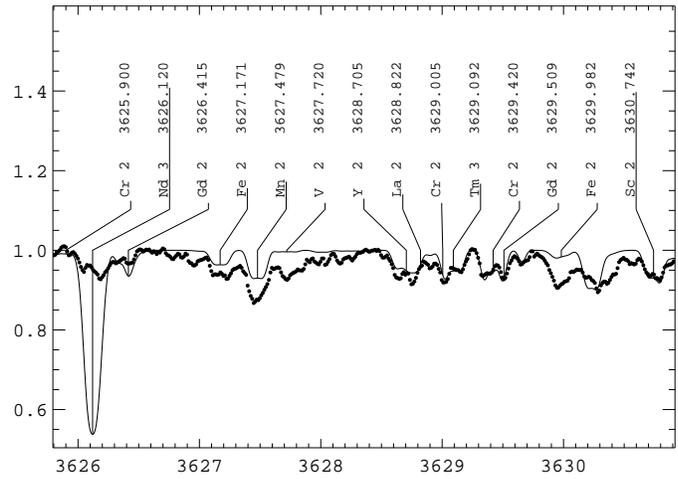}
\caption{The same as in Fig.~\ref{pr-nd-tb} but for Tm\iii\ line. 
No strong feature is observed at 3626.12 \AA\ corresponding to
a strong  Nd\iii\ line predicted by Zhang et al. (\cite{Nd3}) on the basis of Martin et al. (\cite{NBS-REE78})
energy levels.} 
\label{tm3}
\end{figure}

\underline{\bf Thulium}.  Experimental data for Tm\ii\ lines (Wickliffe \& Lawler \cite{Tm2}) and
theoretical calculations for Tm\iii\ lines (Li et al. \cite{Tm3}) were used. The lines of both species
are not strong. Nevertheless, good agreement of the synthesized lines with the observed  Zeeman
splitting confirms reliability of the atomic data (see Fig.\,\ref{tm3} for Tm\iii\ 3629.09~\AA). 

\underline{\bf Ytterbium}. Theoretical oscillator strengths (Bi\'emont et al \cite{Yb3}) were used in
Yb\iii\ analysis. It was possible to model only one line for each Y ion. These lines are weak, but
provide consistent abundance. 

\underline{\bf Lutecium}. The only line which is suitable for the Lu abundance determination is Lu\ii\
6221.89~\AA. It is weak, but in order to fit the line profile we have to account for both the  Zeeman
and \hfs\ splittings. At the same time, derived abundance is not affected by these effects. Oscillator
strength for this line is taken from Den Hartog et al. (\cite{Lu2}) and \hfs\ constants -- from Brix
\& Kopfermann (\cite{BK52}).  Our synthesis confirmed the revised wavelength of this line,
$\lambda$~6221.86~\AA, as derived by Bord et al. (\cite{BCM98}) from the analysis of the solar
spectrum.

Spectral synthesis of the REEs in HD~144897 provides a strong support for the calculated atomic
parameters for the second ions of most of these elements and, hence, for the correct line
classification. Only for Nd\iii\ we obtained inconsistencies between the theoretical calculations and
few rather strong observed features, which disagree either in the Zeeman pattern
(Fig.\,\ref{nd3-3590}) or in the intensity (Fig.\,\ref{tm3}). Taking into account the fact that for
all second ions of REE calculations provide quite reliable atomic data, we may conclude that poor
knowledge of the Nd\iii\ spectrum rather than a problem with the theoretical line profile calculations
is the reason for the observed inconsistencies in the case of Nd\iii\ lines.

\section{Nd\iii\ classification}
\label{nd3}

The Nd\iii\ spectrum is very poorly known. Martin et al. (\cite{NBS-REE78})  compiled 5
levels of the 4f$^4$\,$^5$I ground term and 24 levels of the  4f$^3$5d configuration using
unpublished material by Crosswhite (1976).  The latter list of Nd\iii\ lines was not
published but widely circulates in the  astrophysical community. Recent study of the
Nd\iii\ spectrum was  made by Aldenius (\cite{Ald01}) by means of Fourier Transform
Spectroscopy.  She measured with a precision up to 0.005\cm\ in wavenumbers the 
wavelengths of 58 Nd\iii\ lines, 38 of which were identified with the  levels published by
Martin et al. (\cite{NBS-REE78}). The accuracy of the levels  was significantly improved.
It should be mentioned that Crosswite's and Aldenius'  wavelength lists are partially
inconsistent. Not all lines from  Aldenius' list are present in Crosswhite's one, and some
strong  Crosswhite's lines are absent in Aldenius' list. Theoretical  interpretation of
the Nd\iii\ spectrum was performed by Bord ({\cite{Bord00})  and Zhang et al.
(\cite{Nd3}). In the last paper the experimental  lifetimes of five 4f$^3$\,5d levels were
also reported.

Aldenius (\cite{Ald01}) classified 38 Nd\iii\ lines  and suggested an identification of 20
new Nd\iii\ lines based on the laboratory spectrum, but did not provide their
classification. In our classification work we used 27 the most intensive lines from her
list of 38   classified lines and 9 lines from the list of unclassified ones.  Rejected
are the lines of low intensity (less than 3 in Aldenius'   intensity scale) and the line
4781.014~\AA\ (intensity 10), which is practically invisible  in spectra of Ap and roAp
stars. These spectra provide us with additional wavelength data.  We mentioned in the
Sect.~\ref{intro} that Nd\iii\ lines show outstanding  pulsational characteristics in roAp
stars, which we consider as a necessary (but not sufficient) condition for a spectral
feature to belong to Nd\iii. We chose a set of  unidentified lines in the spectra of roAp
stars $\gamma$~Equ (Kochukhov et al. \cite{gequ})  and HD~24712 (Sachkov et al.
\cite{Rome}), which have pulsational amplitudes and phases similar to those for classified
Nd\iii\ lines, and measured wavelengths of these lines. Because of the large magnetic
field, the spectrum  of HD~144897 is not  suitable for precise wavelength measurements,
therefore we carried out line position determinations using the spectrum of roAp
HD~217522. Moderate magnetic field (\bs$<$2 kG)  and slow rotation (\vs$\le$3\,\kms) of
this star provide an accuracy $\sim$0.005~\AA\  for unblended lines. The spectrum of
HD\,217522 was obtained during the same  observational program 68.D-0254 and was reduced
in the same way as our data on HD~144897. Then we applied a classification procedure to
the whole line sample using the code {\sc iden} (Azarov  \cite{Az93}) for identification
of atomic spectra.  

As in previous studies (Bord \cite{Bord00}; Zhang et al. \cite{Nd3}), our calculations of 
the Nd\iii\ atomic structure were made with the Cowan code (Cowan \cite{cowan81}). For 
the prediction of the effective energy parameters and electrostatic  parameters in the
4f$^4$ and 4f$^3$ shells we took advantage of the generalized  parametric studies of
doubly-ionized lanthanides (Wyart et al. \cite{Pr3}).  Table\,\ref{fit} (Online material)
lists the energy parameters of the 4f$^4$ and 4f$^3$5d  configurations after fitting to
all observed levels. 4f$^4$ + 4f$^2$5d$^2$ + 4f$^3$6p configurations were  included in the
even  system of levels. 4f$^3$5d + 4f$^3$6s + 4f$^2$5d6p configurations form the energy 
matrix of the odd set of levels. All electrostatic parameters  for unknown configurations
as well as the parameters of  configuration interactions were fixed at the 0.85 level
relative to the  corresponding Hartree-Fock values. Average deviation $\sigma$\,=55\,\cm\
of our fitting  is the same as in the previous calculations, but due to  introduction of
the effective energy parameters ($\alpha, \beta, \gamma$  and F$^1$(4f,5d)) and scaling
factors for F$^k$(4f,4f), the Slater integrals predictions for unknown energy levels are
expected to be more  accurate than in previous studies by Bord (\cite{Bord00}) and Zhang
et al. (\cite{Nd3}).

A list of identified Nd\iii\ lines is given in Table\,\ref{class} and found  energy levels
are presented in Table\,\ref{levels} (Online material). First column of  Table\,\ref{class}
contains recommended wavelengths for 71 lines. They coincide  with the Aldenius'
wavelengths where available (column 2). Column 3  shows the wavelengths from the
Crosswhite's list. In the 4th column the wavelengths measured in the spectrum of roAp star
HD~217522 are  given. For the lines absent in  Aldenius' list but originated from the
levels defined at least by one  wavelength from Aldenius' list we recommend the
wavelengths  calculated from the level energies (Ritz values). We also give Ritz  values
for transitions from some levels established by Crosswhite's  wavelengths (where Aldenius'
measurements were not available). In  other cases recommended wavelengths coincide with
the stellar  measurements.

An agreement between  the theoretical Zeeman structure and intensity of the lines from
Table\,\ref{class} and the observed  features in HD~144897 was accepted as a sufficient
condition for correct Nd\iii\ classification.  Mean abundance obtained for 39 newly 
classified lines is $\log ({\rm Nd}/N_{\rm tot})=-6.60\pm0.20$ and it agrees very  well with the
abundance derived from the known lines (Table\,\ref{Abund}).  Figure\,\ref{nd3-4914} gives
an example how newly classified Nd\iii\ lines fit the  observed spectra in the
4900--4915~\AA\ region. The top panel displays this  spectral part in HD 144897, and the
spectrum of roAp star HD 24712  (Sachkov et al. \cite{Rome}) is shown in the bottom panel.
Dashed line in  the bottom panel represents scaled pulsation pattern in roAp stars 
(standard deviation of the time-series spectra from the averaged  spectrum) where
double-peaks show a position of the lines with  pulsation radial velocity variations. Only
one pulsating line, Nd\iii\ 4903.24~\AA,  was previously known and classified, and another
line 4914.09~\AA\ was  listed (but not classified) as Nd\iii\ by Aldenius (\cite{Ald01}). 
Previous classification of the Nd\iii\ 4903.24~\AA\ line does not  correspond to the
observed magnetic splitting (dashed line in the  upper panel of Fig.\,\ref{nd3-4914}). Two
other lines, 4911.65 and 4912.94~\AA, with the same  pulsation signature are also seen in
Fig.\,\ref{nd3-4914}. All 4 lines are  classified in this work and validity of the
identification is proved  by Zeeman calculations in the spectrum of HD~144897 (upper panel
in  Fig.\,\ref{nd3-4914}). A good fit to the stellar spectrum for one more newly
classified Nd\iii\ 5851.542~\AA\ line is shown in Fig.\,\ref{pr-nd-tb}. 


\begin{figure*}[!th]
\includegraphics[width=85mm]{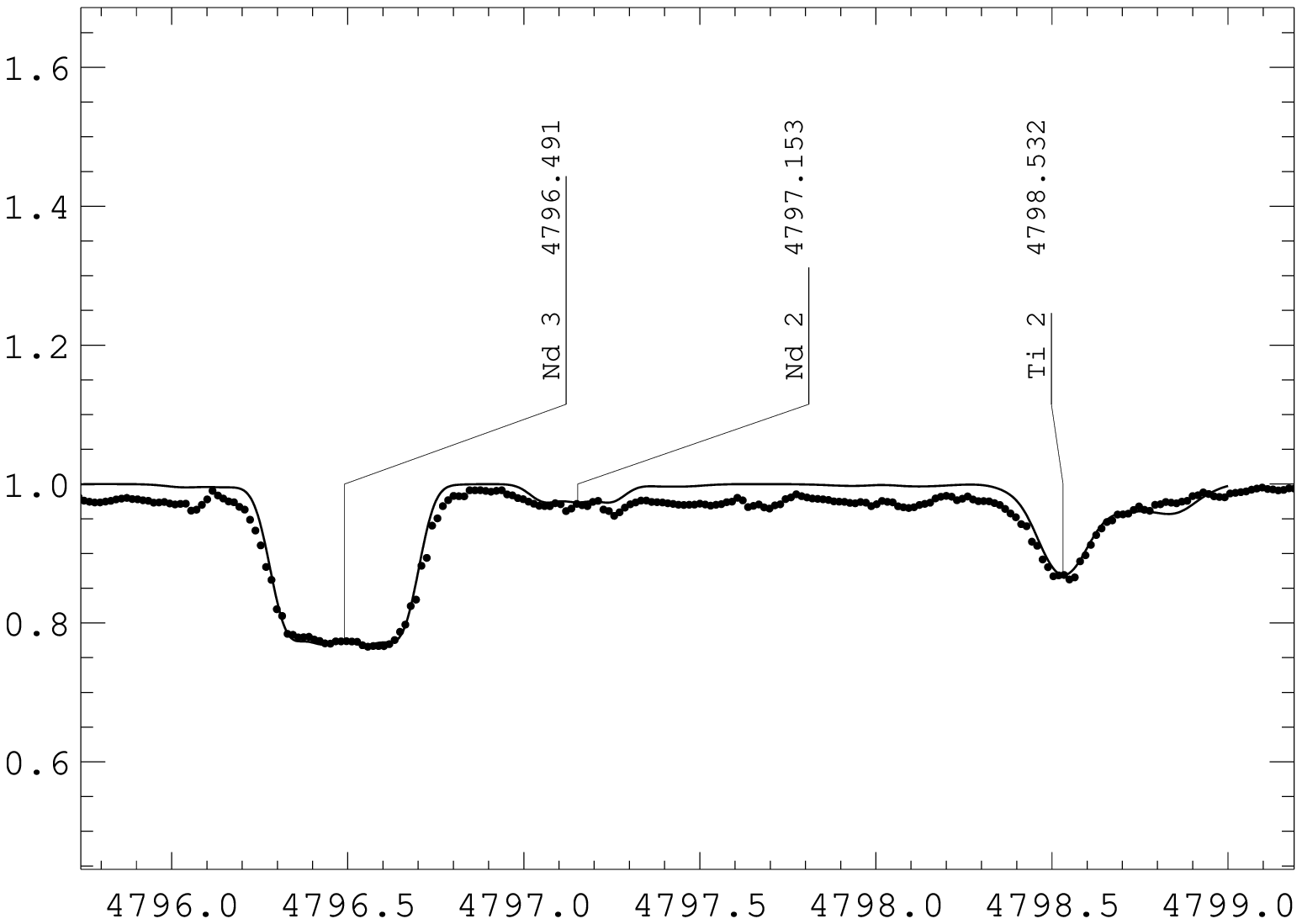}\hspace{0.5cm}\includegraphics[width=85mm]{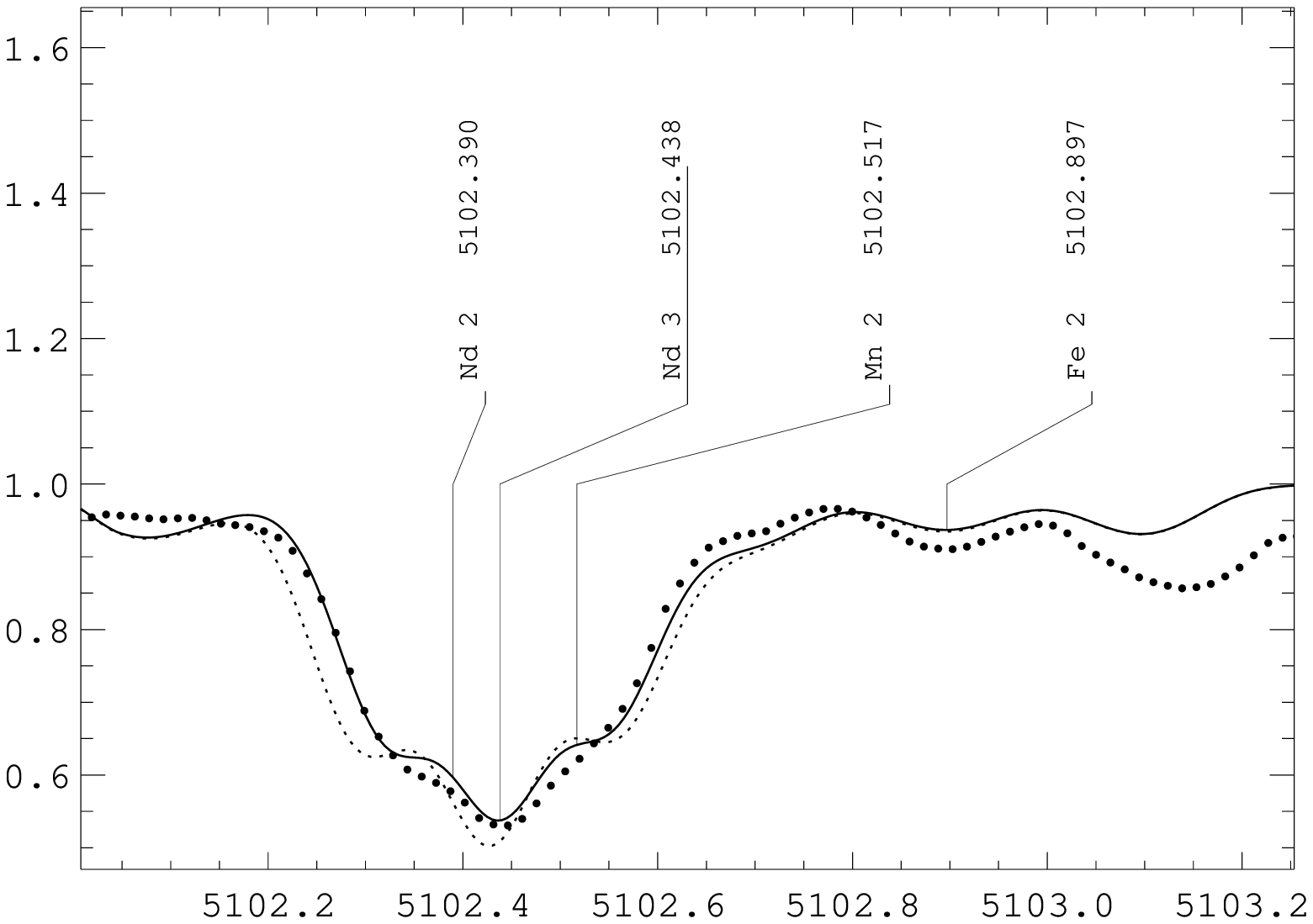}
\caption{The Nd\iii\ lines from the $^4$I$^5$I$_{5,6}$ -- $^4$I$^5$I$_6^o$ transition array. Calculations with the old classification for
$\lambda$~5102.44 line are shown by the dotted line.} 
\label{nd3-4796}
\end{figure*}

\begin{table*}[!th]
\caption{Comparison between the measured and calculated lifetimes of the 
5d\,(4F)$^5$H$_J$ levels  in a framework of new identifications. }
\label{lif}
\begin{center}
\begin{tabular}{ccccc|cccc}
\noalign{\smallskip}
        
\hline
\hline
 &\multicolumn{4}{c|}{Zhang et al. (\cite{Nd3})}& \multicolumn{4}{c}{Present work}\\
  &            &          &\multicolumn{2}{c|}{Lifetime (ns)}&  &  & \multicolumn{2}{c}{Lifetime (ns)}\\   
$J$ & $E$ (\cm) & $\lambda$ (\AA) &exp& theor& $E$ (\cm) & $\lambda$ (\AA) &exp&theor\\
\hline
3 & 27788.2& 3598& 110(10)& 110&   27675.003& 3612& ~70(~5)&  87\\
4 & 28745.3& 3621& 170(10)& 108&   28745.285& 3621& 170(10)&  54\\
5 & 30232.3& 3590& 105(15)& 108&   30175.690& 3598& 110(10)&  69\\
6 & 31394.6& 3612& ~70(~5)& ~81&   31559.191& 3590& 105(15)&  87\\
7 & 32832.6& 3604& ~87(~7)& ~98&   32832.419& 3604& ~87(~7)&  86\\
\hline
\end{tabular}
\end{center}
\end{table*}

Table\,\ref{levels} (Online material) contains the energies of all levels of the 4f$^3$5d 
configuration with $J$ = 3--10 below 33000\cm. The energies of 35  levels were determined. We
were able to confirm the energies only for 11  out of 24 levels compiled by Martin et al.
(\cite{NBS-REE78}). Oscillator strengths of 23 transitions from these 11 levels  to the lower 4f$^4$\,$^5$I term, 
calculated by Zhang et al. (\cite{Nd3}) and in the present work agree within 0.14 dex in average.    
The  energies of almost all levels with $J$ = 4--8 are based now on
the  identification of 2 or 3 lines. The levels with $J$ = 3 are defined by  one line and
need further confirmation. Calculated composition of  the levels (column 4) shows that
some of the levels are highly mixed,  and leading contribution is given by the same LS 
term. To avoid  possible ambiguities, we supplied designations of such levels by 
additional letters A or B in Table\,\ref{class} (see also Table\,\ref{levels} (Online
material)). 

Previous identification was changed for a few  strong lines. For example, the changes were
made for the ($^4$I)$^5$I$^{\rm o}$ J=5,6,8 levels. 

($^4$I)$^5$I$^{\rm o}_5$: New value 20348.715\cm\ is supported by 3 lines, 5203.9236,
4912.946 and 5566.022 \AA, observed in stellar spectra with the  intensities corresponding
to the calculated $gf$ values.  Previous energy of the level 20388.980\cm\  was based on
two lines 5193.0397 and 4903.2410 \AA, but Zeeman structure calculated from this
classification does not fit the observed spectrum of HD~144897. With the new
classification these two lines  perfectly match a transition array from the
($^4$I)$^5$H$^{\rm o}_4$ level (Fig.\,\ref{nd3-4914}).

($^4$I)$^5$I$^{\rm o}_6$: Aldenius (\cite{Ald01}) found only one line 4781.0141 \AA\
(mentioned  above) as the $^5$I$_5$ -- ($^4$I)$^5$I$^{\rm o}_6$ transition to support the
value 22048\cm\  (Martin et al. \cite{NBS-REE78}). Another expected transition to the
$^5$I$^{\rm o}_6$ level at 5084.927 \AA\ with 10 times larger transition  probability was
not measured by Aldenius. A rather strong line at 5085.0 \AA\ is present in  stellar spectra,
but the predicted Zeeman pattern based on Martin et al. energy levels does not fit the observed feature.
There is no
observable feature at 4781.014~\AA\ in any  of our stars (HD~144897, $\gamma$~Equ, HD~24712 or
HD~217522) which may be  compared to rather strong theoretically predicted line based on 
previous identification. Therefore, we reject previous energy for ($^4$I)$^5$I$^{\rm o}_6$
and propose instead two other lines 4796.4999  and 5102.4258 (blended by Nd\ii) for the
transitions from the upper  ($^4$I)$^5$I$^{\rm o}_6$ level to $^5$I$_{5,6}$ lower levels.
Both lines are fitted fairly well with the new transition probabilities and Land\'e
factors  (Fig.\,\ref{nd3-4796}). The identification is supported by the third weak line to
the  $^5$I$_7$ level at 5473.14~\AA. The corrected ($^4$I)$^5$I$^{\rm o}_6$ energy for the
level is 21980.505\cm.

($^4$I)$^5$I$^{\rm o}_6$: For previous value 24686.4\cm\ (Martin et al.
\cite{NBS-REE78})   Aldenius (\cite{Ald01}) found two lines at 5102.4278 and 4766.9971 \AA.
Theoretical calculations (Zhang et al. (\cite{Nd3}) predict strong line at 4766.9971~\AA,
while no feature is observed in stellar spectra at this wavelength. We attributed the
first line to the  ($^4$I)$^5$I$^{\rm o}_6$ level (see above).  The energy 24592.234\cm\
of ($^4$I)$^5$I$^{\rm o}_6$ level was derived  based on the 5127.0441 and 4788.465
\AA\ lines.

Few relatively strong lines from Aldenius' list lines remain unclassified, although for
three of them (4514.128, 4711.333, 5028.520~\AA) the Nd\iii\ identification is supported
by pulsational observations in roAp stars.

\section{Discussion}
\label{disc}

An extended abundance analysis of the magnetic Ap star HD~144897 strongly demonstrated  a
possibility of the adequate fit of the observed spectrum to the theoretical calculations
when atmospheric (temperature, magnetic field, etc.) and atomic line parameters  are
correctly known. In particular, it concerns the rare-earth elements in the first and
second ionization stages, which dominated the spectra of some cool roAp stars, for
example, famous Przybylski's star (HD~101065). Unfortunately, an extended laboratory
analysis  of some REEs in the second ionization stage is not complete. In such cases 
stellar data play an important role in spectrum classification. Usually, the quality of
the theoretical atomic structure calculations are checked by comparison  the observed and
calculated lifetimes. For Nd\iii\ the lifetime measurements were made for 5 levels of the
5d\,(4F)$^5$H  configuration having strong mixture with almost  overlapping 5d\,(4F)$^5$D
levels (Zhang et al. \cite{Nd3}). The mixture results in the presence of two  close lying
levels with the (4F)$^5$H character, which were  distinguished by additional letters A and
B in our Table\,\ref{class}.     Because of the mixture, small variation of the energy
parameters leads to drastic change of the calculated lifetimes. A comparison between the
observed and theoretical lifetimes calculated by Zhang et al. (\cite{Nd3}) and in the
present work is given in Table\,\ref{lif}. It should be mentioned that we suggested new 
energies for the 5d\,(4F)$^5$H levels with $J$ = 3, 5 and 6. Each level with $J$ =  5 or 6
is now based on two lines with appropriate Zeeman structure checked using the HD~144897
spectrum. The $J$ = 3 level is defined by one relatively narrow  line in accordance with
the {\it g}-factor of this level. New classification corresponds better to the observed
feature, although the calculated line is still narrower then the observed one. As a result
we have to  interchange the observed wavelengths and energies  for $J$ = 3, 5 and 6 levels
according to new line classification.

One can notice an apparently better agreement between measured and calculated lifetimes in
Zhang et al. (\cite{Nd3}). However, there is a difficulty in the calculations of the
oscillator strengths for the 5p$^6$4f$^n$ -- 5p$^6$4f$^{n-1}$5d transitions  related to
the fact that the 4f electrons are deeply embedded inside  the 5p$^5$ core as discussed by
Bi\'emont et al. (\cite{dream-Pr3}). As a result, the relativistic Hartree-Fock
calculated lifetimes for the 4f$^{n-1}$5d levels are shorter than the measured ones. 
For example, in Eu\iii\ a ratio of measured  and calculated lifetimes is equal to 3
(Mashonkina et al. \cite{NLTE-Eu}). In  view of this property, the relation between
calculated and measured  lifetimes has an expected tendency in the present work,
but not in the Zhang et al. (\cite{Nd3}) study. 

\section{Conclusions}
\label{concl}

We presented an extensive abundance analysis, with the emphasis on REEs, in the atmosphere
of the tepid magnetic Ap star HD~144897. Thanks to the extremely large wavelength coverage of our
observational material, all stable REEs but La and Lu are studied  using
spectral lines of the first and second ionization stages, thus giving more confidence to
the REE abundance results. The spectrum of HD~144897 is very complex because of the large
overabundance of the Fe-peak elements and  Zeeman splitting of lines in a 8.8~kG magnetic
field. Therefore, a careful abundance and, in particular, stratification analysis of the
most prominent elements, like Cr and Fe, was performed to minimize blending problems in
the REE analysis. Comparison of the atmospheric abundances in HD~144897 and in the other
two tepid Ap stars, HD~170973 (Kato \cite{Kato03}) and HD~116458  (Nishimura et al.
\cite{HR5049}), shows close similarity, in particular, in the REE abundances. On average,
REEs are overabundant by 4 dex compared to the solar abundances, and no serious
violation of odd-even effect is observed. Using a spectrum of the star where clearly
resolved Zeeman patterns for most spectral lines are observed, we were able to check an
accuracy of the experimental (first REE ions) and theoretical (second REE ions) data:
energy level classification, transition probabilities and Land\'e factors. We found that
for all ions but Nd\iii, the profiles of unblended or partially blended REE lines in the
spectrum of HD~144897 are well synthesized adopting for the mean magnetic field modulus the
value 8.8~\,kG. In
contrast, less than half of the previously classified Nd\iii\ lines could be fitted. Using
two criteria for the Nd\iii\ line identification -- peculiar pulsational characteristics
in rapidly oscillating stars and a correspondence between the calculated and observed
Zeeman pattern in magnetic star HD~144897 -- we performed a revision of Nd\iii\
classification. We were able to confirm the energies only for 11  out of 24 levels
compiled by Martin et al. (\cite{NBS-REE78}), and we derived energies for additional 24
levels of the 4f$^3$5d configuration increasing a total number of classified Nd\iii\ lines
with corrected  wavelengths and atomic parameters. 

Our classification of the Nd\iii\ energy levels was partially based on  stellar
measurements. Therefore, precise laboratory measurements in wide spectral
region are required to confirm our results.

\begin{acknowledgements}

We thank Prof. C.R. Cowley for providing us with the unpublished Nd\iii\ line lists by  H.
Crosswhite. TR acknowledges financial support from the RFBR grant 04-02-16788a, the 
Presidium of RAS Program ``Origin and   Evolution of Stars and Galaxies'' and the Austrian
Science Fund (FWF-P17580N2). 

\end{acknowledgements}

\begin{longtable}{llrl|l|rcc|lccl}
\caption{\label{class} Proposed classification for Nd\iii\ lines.}\\
\hline
\hline
\multicolumn{4}{c|}{Wavelength (\AA)} & &\multicolumn{3}{c|}{{Lower level$^a$}} & \multicolumn{4}{c}{Upper level}  \\
{This work$^b$} & {exp$^c$} & {exp$^d$} & {Stellar$^e$} &
$\log(gf)$ & 
$E$ (cm$^{-1}$) & $J$ & {\it g} &
$E$ (cm$^{-1}$) & $J$ & {\it g} & {LS term$^f$} \\
\hline
\endfirsthead
\caption{continued.}\\
\hline\hline
\multicolumn{4}{c|}{Wavelength (\AA)} & &\multicolumn{3}{c|}{{Lower level$^a$}} & \multicolumn{4}{c}{Upper level}  \\
{This work$^b$} & {exp$^c$} & {exp$^d$} & {Stellar$^e$} &
$\log(gf)$ & 
$E$ (cm$^{-1}$) & $J$ & {\it g} &
$E$ (cm$^{-1}$) & $J$ & {\it g} & {LS term$^f$} \\
\hline
\endhead
\hline
\endfoot
\hline
\endfoot
 3427.0017R  &        &     & .00bl & -1.69 &2387.529  & 6.0&  1.071  &31559.191 & 6.0 & 1.151 &($^4$F)$^5$H$^{\rm o}$B\\
 3433.3315   & .3315  & .327& .33bl & -1.70 &3714.537  & 7.0&  1.177  &32832.420 & 7.0 & 1.155 &($^4$F)$^5$H$^{\rm o}$B\\
 3442.7891R  &        & .781& .793  & -1.49 &1137.795  & 5.0&  0.902  &30175.690 & 5.0 & 1.051 &($^4$F)$^5$H$^{\rm o}$B\\
 3476.1892R  &        & .184& .192  & -2.00 &2387.529  & 6.0&  1.071  &31146.429 & 6.0 & 1.060 &($^4$F)$^5$H$^{\rm o}$A\\
 3477.8359   & .8359  & .834& .839  & -1.66 &   0.000  & 4.0&  0.605  &28745.285 & 4.0 & 0.916 &($^4$F)$^5$H$^{\rm o}$ \\
 3537.6116   & .6116  & .607& .63bl & -2.60 &1137.795  & 5.0&  0.902  &29397.378 & 5.0 & 0.982 &($^4$F)$^5$H$^{\rm o}$A\\
 3561.8555R  &        &     & .858  & -2.31 &3714.537  & 7.0&  1.177  &31781.776 & 7.0 & 1.141 &($^4$F)$^5$H$^{\rm o}$A\\
 3590.3291   & .3291  & .332& .330  & -0.59 &3714.537  & 7.0&  1.177  &31559.191 & 6.0 & 1.151 &($^4$F)$^5$H$^{\rm o}$B\\
 3597.6284   & .6284  & .629& .630  & -0.57 &2387.529  & 6.0&  1.071  &30175.690 & 5.0 & 1.051 &($^4$F)$^5$H$^{\rm o}$B\\
 3603.9826   & .9826  &     & .982  & -0.55 &5093.250  & 8.0&  1.247  &32832.420 & 7.0 & 1.155 &($^4$F)$^5$H$^{\rm o}$B\\
 3612.3388   & .3388  &     & .338  & -0.80 &   0.000  & 4.0&  0.605  &27675.003 & 3.0 & 0.836 &($^4$F)$^5$H$^{\rm o}$ \\
 3621.1727   & .1727  & .171& .18bl & -0.52 &1137.795  & 5.0&  0.902  &28745.285 & 4.0 & 0.916 &($^4$F)$^5$H$^{\rm o}$ \\
 3644.3534   & .3534  &     & .357  & -0.85 &3714.537  & 7.0&  1.177  &31146.429 & 6.0 & 1.060 &($^4$F)$^5$H$^{\rm o}$A\\
 3701.3000   & .3000  & .288& .32   & -1.06 &2387.529  & 6.0&  1.071  &29397.378 & 5.0 & 0.982 &($^4$F)$^5$H$^{\rm o}$A\\
 3745.8637   & .8637  &     & .86   & -0.62 &5093.250  & 8.0&  1.247  &31781.776 & 7.0 & 1.141 &($^4$F)$^5$H$^{\rm o}$A\\
 4211.003    &        & .003& .014  & -2.43 &1137.795  & 5.0&  0.902  &24878.420 & 5.0 & 1.141 &($^4$I)$^3$H$^{\rm o}$ \\
 4213.3602R  &        & .356& .367  & -2.66 &3714.537  & 7.0&  1.177  &27441.879 & 8.0 & 1.140 &($^4$I)$^3$K$^{\rm o}$ \\
 4359.485 R  &        &     & .48bl & -3.03 &2387.529  & 6.0&  1.071  &25319.572 & 7.0 & 1.099 &($^4$I)$^3$K$^{\rm o}$ \\
 4414.300    &        & .300& .312  & -2.22 &2387.529  & 6.0&  1.071  &25034.830 & 6.0 & 1.250 &($^4$I)$^5$G$^{\rm o}$ \\
 4444.997 R  &        &     &5.01   & -2.22 &2387.529  & 6.0&  1.071  &24878.420 & 5.0 & 1.141 &($^4$I)$^3$H$^{\rm o}$ \\
 4466.355    &        &     & .355  & -2.52 &3714.537  & 7.0&  1.177  &26097.872 & 7.0 & 1.186 &($^4$I)$^5$H$^{\rm o}$ \\
 4473.2922   & .2922  & .287& .292  & -1.30 &5093.250  & 8.0&  1.247  &27441.879 & 8.0 & 1.140 &($^4$I)$^3$K$^{\rm o}$ \\
 4501.2346R  &        &     & blend & -2.03 &1137.795  & 5.0&  0.902  &23347.692 & 5.0 & 1.153 &($^4$I)$^5$G$^{\rm o}$ \\
 4507.0450R  &        &     & .06bl & -3.24 &   0.000  & 4.0&  0.605  &22181.265 & 5.0 & 1.032 &($^4$I)$^5$H$^{\rm o}$ \\
 4570.660    &        & .660& .643  & -2.06 &1137.795  & 5.0&  0.902  &23010.344 & 4.0 & 1.031 &($^4$I)$^5$G$^{\rm o}$ \\
 4613.1780R  &        &     & .211  & -3.14 &2387.529  & 6.0&  1.071  &24058.489 & 6.0 & 1.122 &($^4$I)$^5$H$^{\rm o}$ \\
 4624.9799   & .9799  & .965& .97bl & -1.88 &2387.529  & 6.0&  1.071  &24003.190 & 7.0 & 1.176 &($^4$I)$^5$I$^{\rm o}$B\\
 4627.254    &        & .254& .254  & -2.08 &3714.537  & 7.0&  1.177  &25319.572 & 7.0 & 1.099 &($^4$I)$^3$K$^{\rm o}$ \\
 4651.6225   & .6225  & .609& .617  & -1.86 &   0.000  & 4.0&  0.605  &21491.858 & 4.0 & 0.948 &($^4$I)$^3$H$^{\rm o}$ \\
 4654.312    &        & .312& .310  & -1.92 &   0.000  & 4.0&  0.605  &21479.437 & 3.0 & 0.825 &($^4$I)$^5$G$^{\rm o}$ \\
 4689.052    &        & .052& .055  & -1.80 &3714.537  & 7.0&  1.177  &25034.830 & 6.0 & 1.250 &($^4$I)$^5$G$^{\rm o}$ \\
 4759.526    &        & .499& .526  & -1.15 &5093.250  & 8.0&  1.247  &26097.872 & 7.0 & 1.186 &($^4$I)$^5$H$^{\rm o}$ \\
 4769.6217   & .6217  & .610& .610  & -1.796&2387.529  & 6.0&  1.071  &23347.692 & 5.0 & 1.153 &($^4$I)$^5$G$^{\rm o}$ \\            
 4788.4617R  &        & .423& .465  & -1.77 &3714.537  & 7.0&  1.177  &24592.234 & 8.0 & 1.229 &($^4$I)$^5$I$^{\rm o}$ \\
 4796.4999   & .4999  & .465& .499  & -1.65 &1137.795  & 5.0&  0.902  &21980.505 & 6.0 & 1.084 &($^4$I)$^5$I$^{\rm o}$ \\
 4821.986    &        & .986& .988  & -2.45 &2387.529  & 6.0&  1.071  &23120.081 & 6.0 & 0.940 &($^4$I)$^3$K$^{\rm o}$ \\
 4903.2410   & .2410  &     & .26bl & -2.21 &   0.000  & 4.0&  0.605  &20388.980 & 4.0 & 0.836 &($^4$I)$^5$H$^{\rm o}$ \\
 4911.6527R  &        &     & .658  & -1.63 &1137.795  & 5.0&  0.902  &21491.858 & 4.0 & 0.948 &($^4$I)$^3$H$^{\rm o}$ \\
 4912.9436R  &        &     & .946  & -1.78 &   0.000  & 4.0&  0.605  &20348.715 & 5.0 & 0.924 &($^4$I)$^5$I$^{\rm o}$ \\
 4914.0941   & .0941  &     & .095  & -1.10 &3714.537  & 7.0&  1.177  &24058.489 & 6.0 & 1.122 &($^4$I)$^5$H$^{\rm o}$ \\
 4921.0431   & .0431  &     & .04bl & -1.89 &2387.529  & 6.0&  1.071  &22702.749 & 7.0 & 1.146 &($^4$I)$^5$I$^{\rm o}$A\\
 4927.4877   & .4877  & .484& .484  & -0.80 &3714.537  & 7.0&  1.177  &24003.190 & 7.0 & 1.176 &($^4$I)$^5$I$^{\rm o}$B\\
 4942.673    &        & .673& .658  & -1.23 &5093.250  & 8.0&  1.247  &25319.572 & 7.0 & 1.099 &($^4$I)$^3$K$^{\rm o}$ \\
 5050.6952   & .6952  &     & .700  & -1.06 &2387.529  & 6.0&  1.071  &22181.265 & 5.0 & 1.032 &($^4$I)$^5$H$^{\rm o}$ \\
 5084.6597R  &        &     & .660  & -2.58 &1137.795  & 5.0&  0.902  &20799.314 & 6.0 & 1.021 &($^4$I)$^3$I$^{\rm o}$ \\
 5102.4278   & .4278  & .430& .437  & -0.62 &2387.529  & 6.0&  1.071  &21980.505 & 6.0 & 1.084 &($^4$I)$^5$I$^{\rm o}$ \\
 5127.0441   & .0441  & .009& .049  & -0.40 &5093.250  & 8.0&  1.247  &24592.234 & 8.0 & 1.229 &($^4$I)$^5$I$^{\rm o}$ \\
 5151.731 R  &        &     & .742  & -1.54 &3714.537  & 7.0&  1.177  &23120.081 & 6.0 & 0.940 &($^4$I)$^3$K$^{\rm o}$ \\
 5152.292    &        & .292& .25bl & -1.21 &   0.000  & 4.0&  0.605  &19403.433 & 3.0 & 0.575 &($^4$I)$^5$H$^{\rm o}$ \\
 5193.0397   & .0397  & .061& .038  & -1.18 &1137.795  & 5.0&  0.902  &20388.980 & 4.0 & 0.836 &($^4$I)$^5$H$^{\rm o}$ \\
 5203.9236   & .9236  & .906& .922  & -0.66 &1137.795  & 5.0&  0.902  &20348.715 & 5.0 & 0.924 &($^4$I)$^5$I$^{\rm o}$ \\
 5264.9604   & .9604  & .959& .965  & -0.72 &3714.537  & 7.0&  1.177  &22702.749 & 7.0 & 1.146 &($^4$I)$^5$I$^{\rm o}$A\\
 5286.7534   & .7534  &     & .728  & -1.90 &5093.250  & 8.0&  1.247  &24003.190 & 7.0 & 1.176 &($^4$I)$^5$I$^{\rm o}$B\\
 5294.1133   & .1133  & .106& .108  & -0.69 &   0.000  & 4.0&  0.605  &18883.652 & 4.0 & 0.640 &($^4$I)$^5$I$^{\rm o}$ \\
 5410.0994   & .0994  &     & .098  & -1.52 &1137.795  & 5.0&  0.902  &19616.608 & 5.0 & 0.869 &($^4$I)$^3$I$^{\rm o}$ \\
 5429.7944   & .7944  &     & .793  & -1.24 &2387.529  & 6.0&  1.071  &20799.314 & 6.0 & 1.021 &($^4$I)$^3$I$^{\rm o}$ \\
 5473.1411R  &        &     & .14bl & -3.03 &3714.537  & 7.0&  1.177  &21980.505 & 6.0 & 1.084 &($^4$I)$^5$I$^{\rm o}$ \\
 5566.0154R  &        &     & .022  & -2.33 &2387.529  & 6.0&  1.071  &20348.715 & 5.0 & 0.924 &($^4$I)$^5$I$^{\rm o}$ \\
 5633.5540   & .5540  &     & .558  & -2.22 &1137.795  & 5.0&  0.902  &18883.652 & 4.0 & 0.640 &($^4$I)$^5$I$^{\rm o}$ \\
 5677.1788   & .1788  &     & .189  & -1.45 &5093.250  & 8.0&  1.247  &22702.749 & 7.0 & 1.146 &($^4$I)$^5$I$^{\rm o}$A\\
 5802.5319R  &        &     & .536  & -1.71 &2387.529  & 6.0&  1.071  &19616.608 & 5.0 & 0.869 &($^4$I)$^3$I$^{\rm o}$ \\
 5845.0201   & .0201  & 5.00& .018  & -1.18 &5093.250  & 8.0&  1.247  &22197.090 & 9.0 & 1.218 &($^4$I)$^5$K$^{\rm o}$ \\
 5851.5419R  &        &     & .533  & -1.55 &3714.537  & 7.0&  1.177  &20799.314 & 6.0 & 1.021 &($^4$I)$^3$I$^{\rm o}$ \\
 5987.6828   & .6828  &     & .679  & -1.26 &3714.537  & 7.0&  1.177  &20410.864 & 8.0 & 1.150 &($^4$I)$^5$K$^{\rm o}$ \\
 6145.0677   & .0677  & .021& .058  & -1.33 &2387.529  & 6.0&  1.071  &18656.240 & 7.0 & 1.054 &($^4$I)$^5$K$^{\rm o}$ \\
 6327.2649   & .2649  & .220& .266  & -1.41 &1137.795  & 5.0&  0.902  &16938.043 & 6.0 & 0.912 &($^4$I)$^5$K$^{\rm o}$ \\
 6526.6288R  &        &     & .638  & -2.51 &5093.250  & 8.0&  1.247  &20410.864 & 8.0 & 1.150 &($^4$I)$^5$K$^{\rm o}$ \\
 6550.2314   & .2314  & .212& .237  & -1.49 &   0.000  & 4.0&  0.605  &15262.420 & 5.0 & 0.688 &($^4$I)$^5$K$^{\rm o}$ \\
 6690.8302R  &        &     & .833  & -2.46 &3714.537  & 7.0&  1.177  &18656.240 & 7.0 & 1.054 &($^4$I)$^5$K$^{\rm o}$ \\
 6870.7137R  &        &     & .71bl & -2.58 &2387.529  & 6.0&  1.071  &16938.043 & 6.0 & 0.912 &($^4$I)$^5$K$^{\rm o}$ \\
 7077.8825R  &        &     & .885  & -2.85 &1137.795  & 5.0&  0.902  &15262.420 & 5.0 & 0.688 &($^4$I)$^5$K$^{\rm o}$ \\
\hline														       
\multicolumn{12}{l}{}\\
\multicolumn{12}{l}{$^a$ \footnotesize{Lower levels belong to the 4f$^4$\,$^5$I term}}\\
\multicolumn{12}{l}{$^b$ \footnotesize{R - wavelengths calculated from known level energies (Ritz wavelengths)}}\\
\multicolumn{12}{l}{$^c$ \footnotesize{Aldenius (\cite{Ald01})}}\\
\multicolumn{12}{l}{$^d$ \footnotesize{Based on data provided by Crosswhite (1976)}}\\
\multicolumn{12}{l}{$^e$ \footnotesize{Measured in HD~217522 spectrum}}\\
\multicolumn{12}{l}{$^f$ \footnotesize{Upper levels belong to the 4f$^3$5d configuration. See text for A,B}}\\
\multicolumn{12}{l}{}\\
\end{longtable}

\Online

\begin{footnotesize}
\begin{longtable}{ccrl}
\caption{\label{lines} Line-by-line abundances in HD~144897.}\\
\hline 
\hline
Wavelength (\AA)& $E_{\rm low}$ (eV) & $\log(gf)$ &$\log (N/N_{\rm tot})$ \\
\hline
\endfirsthead
\caption{Continued.}\\
\hline
\hline
Wavelength (\AA)& $E_{\rm low}$ (eV) & $\log(gf)$ &$\log (N/N_{\rm tot})$ \\
\hline
\endhead
\hline
\endfoot
\multicolumn{4}{l}{\ion{~O}{i}}    \\ 						       
 7771.9413&  9.146&  0.369& -4.00  \\
 7774.1607&  9.146&  0.223& -4.00  \\
 7775.3904&  9.146&  0.001& -4.00  \\
\multicolumn{4}{l}{\ion{Co}{i}}    \\
 3412.3330&  0.514&  0.030& -4.80  \\
 3502.2780&  0.432&  0.070& -4.60  \\
 3506.3120&  0.514& -0.040& -4.60  \\
\multicolumn{4}{l}{\ion{Co}{ii}}   \\
 3415.7720&  2.203& -1.640& -5.10  \\
 3514.2140&  2.274& -2.970& -4.20  \\
 3517.9900&  3.009& -2.560& -4.30  \\
 3523.5460&  2.729& -1.540& -4.80  \\
 3550.7120&  2.985& -2.680& -4.40  \\
 3566.9590&  2.729& -2.440& -4.70  \\
 4915.4230&  3.406& -3.425& -4.20  \\
\multicolumn{4}{l}{\ion{Sr}{ii}}   \\ 
 3380.7070&  2.940&  0.199& -7.10  \\ 
 4161.7920&  2.940& -0.502& -7.00: \\
 4215.5190&  0.000& -0.145& -8.20  \\
 4305.4430&  3.040& -0.136& -7.50  \\
\multicolumn{4}{l}{\ion{~Y}{ii}}   \\
 3242.2800&  0.180&  0.210& -7.75  \\
 4900.1200&  1.033& -0.090& -7.60  \\
 5662.9250&  1.944&  0.160& -7.75  \\
\multicolumn{4}{l}{\ion{Zr}{ii}}   \\
 3129.1530&  0.527& -0.320& -7.30  \\ 
 3129.7630&  0.039& -0.650& -7.30  \\ 
 3138.6830&  0.095& -0.460& -7.30  \\ 
 3241.0420&  0.039& -0.504& -7.60  \\ 
 3305.1530&  0.039& -0.590& -7.20  \\ 
 3344.7860&  1.011& -0.220& -7.40  \\ 
 3354.3900&  0.758& -0.744& -7.30  \\ 
 4149.2170&  0.802& -0.030& -7.60  \\ 
\multicolumn{4}{l}{\ion{Ba}{ii}}   \\ 
 4554.0290&  0.000&  0.170& -9.00  \\
\multicolumn{4}{l}{\ion{La}{ii}}   \\  
 4123.2180&  0.321&  0.130& -7.30  \\
 4655.4800&  1.946&  0.115& -7.40  \\
 4748.7260&  0.927& -0.540& -7.40  \\
 4899.9150&  0.000& -0.730& -7.60  \\
\multicolumn{4}{l}{\ion{Ce}{ii}}   \\ 
 4127.3640&  0.684&  0.350& -6.80  \\
 4142.3970&  0.696&  0.300& -6.80  \\
 4248.6710&  0.684&  0.140& -6.40  \\
 4460.2070&  0.478&  0.320& -6.90  \\
 4560.2800&  0.910&  0.310& -6.80  \\ 
 4562.3590&  0.478&  0.230& -6.80  \\
 4606.4000&  0.910& -0.020& -6.40  \\
 4680.1190&  1.058& -0.430& -6.40  \\
 4737.2710&  1.090& -0.040& -6.80  \\
 5079.6820&  1.384&  0.420& -6.80  \\
\multicolumn{4}{l}{\ion{Ce}{iii}}  \\
 3085.0990&  2.385&  0.010& -6.80  \\
 3106.9800&  2.413& -0.200& -6.50  \\
 3353.2860&  2.663&  0.180& -6.80  \\
 3395.7700&  2.709& -0.500& -6.40  \\ 
 3427.3580&  2.385& -0.170& -6.75  \\
 3454.3870&  2.413& -0.060& -6.70  \\
 3470.9200&  2.413&  0.140& -6.80  \\
 4535.7260&  2.663& -1.600& -6.40  \\
\multicolumn{4}{l}{\ion{Pr}{ii}}   \\
 4206.7190&  0.550&  0.480& -6.70  \\
 5322.7720&  0.483& -0.315& -6.50  \\  
\multicolumn{4}{l}{\ion{Pr}{iii}}  \\
 4725.5590&  2.078& -1.365& -6.70  \\	     
 4929.1150&  0.359& -2.068& -6.70  \\
 5284.6930&  0.173& -0.771& -6.70  \\
 5299.9930&  0.359& -0.520& -6.50  \\
 5844.4080&  1.244& -1.011& -6.70  \\
 5998.9300&  0.173& -1.872& -6.70  \\
 6090.0100&  0.359& -0.871& -6.70  \\
 6160.2330&  0.173& -1.020& -6.70  \\
 6195.6190&  0.000& -1.071& -6.70  \\
 6692.2467&  1.162& -2.111& -6.50  \\
 7030.3850&  0.359& -0.929& -6.50  \\
 7076.6120&  0.173& -1.429& -6.50  \\
 7781.9830&  0.000& -1.276& -7.00  \\
 7888.1236&  1.346& -1.267& -7.00  \\
\multicolumn{4}{l}{\ion{Nd}{ii}}  \\				
 4061.0800&  0.471&  0.550& -6.80\\
 4156.0780&  0.182&  0.160& -6.50\\
 4462.9790&  0.559&  0.040& -6.50\\
 4542.6000&  0.742& -0.280& -6.50\\
 4645.7600&  0.559& -0.760& -6.40\\
 4706.5430&  0.000& -0.710& -6.50\\
 4715.5860&  0.205& -0.900& -6.50\\
 4724.3570&  0.742& -0.430& -6.50\\
 5319.8150&  0.550& -0.140& -6.40\\
\multicolumn{4}{l}{\ion{Nd}{iii}: previous classification}  \\
 4927.4880&  0.461& -0.800& -6.75\\
 5286.7530&  0.632& -1.880& -6.25\\
 5294.1130&  0.000& -0.700& -6.55\\
 5264.9600&  0.461& -0.720& -6.60\\
 5633.5540&  0.141& -2.220& -6.60\\
 5677.1790&  0.632& -1.450& -6.50\\ 
 5845.0200&  0.632& -1.180& -6.50\\ 
 5987.6830&  0.461& -1.260& -6.60\\
 6145.0680&  0.296& -1.340& -6.50\\
 6327.2650&  0.141& -1.410& -6.35\\
 6550.2310&  0.000& -1.490& -6.50\\
 6690.8300&  0.461& -2.440& -6.10\\
 7077.8830&  0.141& -2.840& -6.10\\
\multicolumn{4}{l}{\ion{Nd}{iii}: new and reclassified  lines} \\
 3442.7890&  0.141& -1.490& -6.70\\
 3476.1890&  0.296& -2.000& -6.50\\
 3477.8360&  0.000& -1.660& -6.60\\
 3537.6120&  0.141& -2.600& -6.10\\ 
 3590.3290&  0.461& -0.590& -6.85\\
 3597.6280&  0.296& -0.570& -6.50\\
 3612.3388&  0.000& -0.800& -7.20\\
 3644.3530&  0.460& -0.850& -6.60\\
 4414.2978&  0.296& -2.220& -6.50\\
 4445.9970&  0.296& -2.220& -6.60\\ 
 4466.3550&  0.460& -2.520& -6.30\\
 4473.2920&  0.631& -1.300& -6.90\\  
 4570.6430&  0.141& -2.060& -6.65\\
 4613.1780&  0.296& -3.140& -6.3:\\
 4627.2540&  0.460& -2.080& -6.55\\
 4651.6230&  0.141& -1.860& -6.80\\
 4654.3115&  0.000& -1.920& -6.80\\ 
 4689.0552&  0.461& -1.800& -6.65\\
 4759.5260&  0.631& -1.150& -6.70\\
 4769.6217&  0.296& -1.800& -6.70\\
 4788.4620&  0.460& -1.770& -6.70\\
 4796.5000&  0.141& -1.650& -6.70\\ 
 4821.9860&  0.296& -2.450& -6.40\\
 4903.2410&  0.000& -2.210& -6.65\\
 4911.6530&  0.141& -1.630& -6.75\\
 4912.9440&  0.000& -1.780& -6.70\\
 4914.0940&  0.460& -1.100& -6.70\\
 4942.6730&  0.631& -1.230& -6.70\\
 5050.6920&  0.296& -1.060& -6.30\\
 5084.6590&  0.141& -2.580& -6.50\\
 5102.4380&  0.296& -0.620& -6.20\\
 5127.0440&  0.631& -0.400& -6.50\\ 
 5151.7310&  0.460& -1.540& -6.65\\
 5193.0400&  0.141& -1.180& -6.70\\
 5203.9240&  0.141& -0.660& -6.70\\
 5410.0990&  0.141& -1.520& -6.50\\
 5429.7940&  0.296& -1.240& -6.60\\
 5566.0150&  0.296& -2.330& -6.40\\
 5851.5420&  0.460& -1.550& -6.60\\  
\multicolumn{4}{l}{\ion{Sm}{ii}} \\
 3568.2710&  0.485&  0.290& -7.20\\	   
 4420.5240&  0.333& -0.430& -6.75\\
 4424.3370&  0.485&  0.140& -6.86\\
 4467.3410&  0.659&  0.150& -7.10\\
\multicolumn{4}{l}{\ion{Sm}{iii}}\\
 3269.3290&  0.388& -0.960& -7.00\\
 3355.3370&  0.388& -1.510& -7.00\\
 3392.2610&  0.100& -1.990& -7.05\\
 3414.4750&  0.282& -1.370& -7.00\\
 3433.5990&  0.185& -1.360& -6.95\\
 3453.2090&  0.100& -1.520& -7.00\\
 3491.3030&  0.000& -2.960& -6.55\\
 3527.4520&  0.036& -1.920& -7.20\\
 3528.7730&  0.282& -2.040& -6.60\\
 3536.1650&  0.185& -1.900& -6.80\\
\multicolumn{4}{l}{\ion{Gd}{ii}}  \\
 3330.3390&  0.991&  0.119& -7.00 \\
 3360.7120&  0.032& -0.240& -7.10 \\
 3362.2390&  0.079&  0.294& -7.10 \\
 4049.4230&  0.662& -0.124& -7.10 \\
 4049.8550&  0.991&  0.429& -7.20 \\
 4130.3660&  0.731& -0.090& -7.00 \\
 4131.4730&  1.406& -0.105& -7.00 \\
 4162.7330&  0.492& -0.720& -7.10 \\
 4163.0870&  0.662& -0.837& -7.00 \\
 4184.2580&  0.492& -0.080& -6.70 \\
 4215.0220&  0.427& -0.550& -6.70 \\
 4732.6090&  1.102& -0.695& -6.70 \\
 4755.3430&  2.449& -0.235& -6.70 \\
\multicolumn{4}{l}{\ion{Gd}{iii}} \\
 3118.0410&  1.432&  0.000& -6.70 \\
 3176.6610&  1.432& -0.540& -6.50 \\
\multicolumn{4}{l}{\ion{Tb}{ii}}  \\
 3509.1440&  0.000&  0.700& -7.80 \\
 3567.3490&  0.439&  0.190& -7.80 \\
 3568.5100&  0.000&  0.360& -7.90 \\
\multicolumn{4}{l}{\ion{Tb}{iii}} \\				     
 4774.1209&  0.000& -1.412& -8.20 \\
 5505.4080&  0.000& -0.815& -8.10 \\
 5847.2310&  0.348& -1.013& -8.00 \\
 6092.8960&  0.587& -1.133& -7.80 \\
 6323.6190&  0.776& -1.209& -7.80 \\
 6687.6980&  1.027& -1.355& -7.60 \\
\multicolumn{4}{l}{\ion{Dy}{ii}}  \\
 3215.1920&  1.662&  0.780& -7.20 \\
 3407.7960&  0.000&  0.150& -7.35 \\
 3434.3690&  0.000& -0.430& -6.60:\\
 3460.9690&  0.000& -0.160& -7.00 \\
 3487.6110&  2.500&  0.980& -7.00 \\
 3523.9830&  0.538&  0.430& -7.10 \\
 3531.7070&  0.000&  0.790& -7.30 \\
 3538.5190&  0.000& -0.190& -7.00 \\
 3550.2180&  0.590&  0.463& -7.30 \\
 3574.1530&  0.925&  0.284& -7.30 \\
 3595.0340&  1.314&  0.760& -7.20 \\
\multicolumn{4}{l}{\ion{Eu}{ii}}  \\
 4129.7250&  0.000&  0.173& -8.00 \\
 4205.0420&  0.000&  0.120& -7.80 \\
 6437.6400&  1.320& -0.273& -7.50 \\
 6645.1200&  1.380&  0.205& -7.40 \\
\multicolumn{4}{l}{\ion{Eu}{iii}} \\ 
 6666.3470&  3.977& -1.470& -6.40 \\ 
 6772.2399&  4.005& -2.840& -6.00 \\ 
 6976.0288&  3.496& -2.280& -6.3: \\ 
 7221.8380&  4.006& -1.560& -6.60 \\ 
 7750.5974&  4.316& -2.170& -6.10 \\ 
 8379.1830&  3.961& -1.850& -6.50 \\ 
\multicolumn{4}{l}{\ion{Dy}{iii}}\\
 3078.0560&  4.098& -0.490& -6.50\\
 3085.9810&  3.836& -0.600& -6.50\\
 3369.6550&  0.000& -1.110& -7.20\\ 
 4401.5670&  0.881& -1.430& -7.40\\
 4510.0270&  0.881& -1.850& -7.15\\
 4572.8860&  0.000& -1.720& -7.20\\
\multicolumn{4}{l}{\ion{Ho}{ii}} \\
 3416.4440&  0.079&  0.260& -7.90\\
 3456.0100&  0.000&  0.760& -8.10\\
 3484.8300&  0.079&  0.280& -8.00\\
\multicolumn{4}{l}{\ion{Ho}{iii}} \\
 3435.2690&  1.072& -0.970& -7.80\\
 3581.4470&  0.000& -0.890& -8.00\\ 
 4267.0650&  0.674& -1.440& -8.20\\
 4416.1360&  0.000& -1.550& -8.15\\
 4448.1040&  0.674& -1.180& -8.15\\
 4494.5230&  0.000& -1.360& -8.20\\
\multicolumn{4}{l}{\ion{Er}{ii}} \\
 3303.9530&  1.701&  0.723& -7.45\\
 3499.1030&  0.055&  0.139& -7.65\\
\multicolumn{4}{l}{\ion{Er}{iii}} \\  
 3070.4020&  0.000& -1.830& -7.25 \\
 3073.5360&  0.630& -1.640& -7.10 \\
 3469.0070&  0.000& -1.660& -7.30:\\
 4422.3140&  0.000& -1.740& -7.40 \\
 4540.7120&  0.000& -2.540& -7.00 \\
 4735.5540&  0.630& -1.580& -7.30 \\
 4783.1150&  0.864& -2.010& -7.10 \\
\multicolumn{4}{l}{\ion{Tm}{ii}} \\
 3267.3970&  1.111&  0.454& -7.90 \\
 3309.8010&  1.111&  0.423&$\le$-8.3\\
 3462.1970&  0.000&  0.030& -8.15 \\
\multicolumn{4}{l}{\ion{Tm}{iii}} \\
 3098.6370&  0.000& -2.250& -7.50\\
 3273.9130&  0.000& -1.460& -7.90\\
 3629.0920&  0.000& -1.960& -7.70\\ 
\multicolumn{4}{l}{\ion{Yb}{ii}} \\
 5335.1590&  3.789& -0.260& -7.50:\\
\multicolumn{4}{l}{\ion{Yb}{iii}} \\
 3092.4960&  4.979& -0.370& -7.50\\
\multicolumn{4}{l}{\ion{Lu}{ii}} \\
 6221.8600&  1.541& -0.760& -8.60\\ 
\end{longtable}   
\end{footnotesize}

\begin{footnotesize}
\begin{center}
\begin{longtable}{lcccc}
	\caption{\label{fit} Fitted (FIT) and Hartree-Fock (HF) energy parameters (\cm)
            and their ratios for the 4f$^4$ and 4f$^3$5d configurations of Nd\iii.  }\\
	\noalign{\smallskip}
\hline
\hline
Parameter &     FIT   & HF    & ratio & Remark \\
\hline
$E_{av}$(4f$^4$)   &   ~31510(~~0) & ~41429&       & \\
 F$^2$(4f,4f)      &   ~66417~~~~~ & ~92632& 0.717 &f~\\
 F$^4$(4f,4f)      &   ~45796~~~~~ & ~57678& 0.794 &f~\\
 F$^6$(4f,4f)      &   ~31773~~~~~ & ~41370& 0.768 &f~\\
$\alpha$           &	  ~~~~30~~~~~ &       &       &f~\\
$\beta$            &	  ~~-823~~~~~ &       &       &f~\\
$\gamma$           &	  ~~1268~~~~~ &       &       &f~\\ 
$\zeta$(4f)        &	  ~~~774(~~0) & ~~~849& 0.912 & \\
$\sigma$           &	  ~~~~~1~~~~~ &       &       & \\
\hline    
$E_{av}$(4f$^3$5d) &	  ~44696(~63) & ~42684&       & \\
 F$^2$(4f,4f)	   &	  ~75830(211) & 101374& 0.748 &r1\\
 F$^4$(4f,4f)	   &	  ~52641(146) & ~63549& 0.828 &r1\\
 F$^6$(4f,4f)	   &	  ~36618(102) & ~45703& 0.801 &r1\\
$\alpha$           &      ~~~~30~~~~~ &	      &	      &f~\\
$\beta$            &      ~~-823~~~~~ &	      &	      &f~\\
$\gamma$           &      ~~1268~~~~~ &	      &	      &f~\\
$\zeta$(4f)        &	  ~~~890(~~5) & ~~~946& 0.940 & \\
$\zeta$(5d)        &	  ~~~815(~14) & ~~~830& 0.981 & \\
 F$^1$(4f,5d)      &   ~~1200~~~~~ &	   &	   &f~\\
 F$^2$(4f,5d)      &   ~20656(363) & ~26407& 0.782 &r2\\
 F$^4$(4f,5d)      &   ~10185(179) & ~13021& 0.782 &r2\\
 G$^1$(4f,5d)      &   ~~9168(~82) & ~12823& 0.715 &r3\\
 G$^3$(4f,5d)      &   ~~7294(~65) & ~10202& 0.715 &r3\\
 G$^5$(4f,5d)      &   ~~5530(~50) & ~~7734& 0.715 &r3\\
$\sigma$          &	  ~~~~55~~~~~ &       &       & \\
\hline	  
\end{longtable}
\end{center}												      
f -- parameter is fixed \\
r1 -- parameter ratios are fixed on the same values as in 4f$^4$ \\
r2, r3 -- parameter ratios are fixed on HF values.
	\end{footnotesize}											      
	\begin{footnotesize}
	\begin{center}
\begin{longtable}{l|r|c|l|c}
	\caption{\label{levels} Energy levels (\cm) of the 4f$^3$5d ($J$=3-10) configuration 
           of Nd\iii\ below 33000\cm. }\\
	\noalign{\smallskip}
\hline
\hline
$E^{\rm a}$ &$\Delta$$E^{\rm b}$ & $g$ &Composition$^{\rm c}$ &Remark$^{\rm d}$\\ 
\hline
\multicolumn{5}{l}{$J=3$}\\       
\hline
    19403.433& -183 & 0.575 &	67\% ( $^4$I)$^5$H~ +18\% ( $^4$I)$^3$G +~6\% ( $^4$F)$^5$H &\\   
    21479.437&   48 & 0.825 &	60\% ( $^4$I)$^5$G~ +18\% ( $^4$I)$^3$G +15\% ( $^4$I)$^5$H &\\  
   (24094)   &      & 0.789 &	56\% ( $^4$I)$^3$G~ +30\% ( $^4$I)$^5$G +~7\% ( $^4$I)$^5$H &\\   
    27675.003&   45 & 0.836 &	51\% ( $^4$F)$^5$H~ +10\% ( $^4$F)$^5$D +~9\% ( $^4$S)$^5$D &\\   
   (27851)   &      & 1.072 &	33\% ( $^4$F)$^5$H~ +20\% ( $^4$F)$^5$D +17\% ( $^4$S)$^5$D &\\   
   (29436)   &      & 0.952 &	81\% ( $^4$F)$^5$G~ +~7\% ( $^4$F)$^3$F +~2\% ($^2_1$D)$^3$F \\   
   (30765)   &      & 1.219 &	14\% ( $^4$S)$^5$D~ +11\% ($^2_2$H)$^1$F +11\% ($^2_2$H)$^3$F \\   
   (31159)   &      & 1.160 &	22\% ( $^4$F)$^3$F~ +18\% ( $^4$G)$^3$F +17\% ( $^4$F)$^5$P &\\  
   (31572)   &      & 1.313 &	31\% ( $^4$F)$^5$P~ +22\% ( $^4$F)$^3$D +~9\% ( $^4$S)$^3$D &\\   
   (32438)   &      & 1.233 &	47\% ( $^4$F)$^5$F~ +10\% ( $^4$S)$^3$D +~7\% ( $^4$F)$^5$P &\\   
   (32543)   &      & 1.217 &	29\% ( $^4$F)$^5$F~ +24\% ( $^4$F)$^5$P +10\% ($^2_2$H)$^3$G&\\   
\hline											    
\multicolumn{5}{l}{$J=4$}       \\
\hline
    18883.652&   74 & 0.640 &	81\% ( $^4$I)$^5$I~ +15\% ( $^4$I)$^3$H +~1\% ( $^4$I)$^5$H &\\   
    20388.980& - 11 & 0.836 &	46\% ( $^4$I)$^5$H~ +28\% ( $^4$I)$^3$H +11\% ( $^4$I)$^5$I &\\   
    21491.858&   26 & 0.948 &	19\% ( $^4$I)$^3$H~ +29\% ( $^4$I)$^5$H +20\% ( $^4$I)$^5$G &\\     
    23010.343&   36 & 1.031 &	56\% ( $^4$I)$^5$G~ +18\% ( $^4$I)$^3$H +11\% ( $^4$I)$^5$H &\\   
   (26481)   &      & 1.057 &	71\% ( $^4$I)$^3$G~ +17\% ( $^4$I)$^5$G +~4\% ( $^4$I)$^5$H &\\   
    28745.285& - 43 & 0.916 &	83\% ( $^4$F)$^5$H~ +~7\% ( $^4$I)$^5$H +~2\% ($^2_1$D)$^3$G &\\   
   (29113)   &      & 1.421 &	38\% ( $^4$F)$^5$D~ +32\% ( $^4$S)$^5$D +~7\% ( $^4$G)$^5$D  &\\   
   (30252)   &      & 1.075 &	62\% ( $^4$F)$^5$G~ +10\% ( $^4$F)$^3$H +~5\% ($^2_1$G)$^3$H &\\   
   (31060)   &      & 0.928 &	21\% ($^2_2$H)$^3$H~ +18\% ( $^4$F)$^5$G +18\% ( $^4$F)$^3$H &\\   
   (31932)   &      & 0.969 &	15\% ($^2_2$H)1G~ +14\% ($^2_2$H)$^3$G +~9\% ( $^4$F)$^3$H   & \\	 
   (32640)   &      & 0.719 &	69\% ( $^4$G)$^5$I~ +~4\% ($^2_2$H)$^3$G +~3\% ($^2_2$H)$^3$H& \\   
\hline
\multicolumn{5}{l}{$J=5$}                                              \\
\hline
    15262.420&   28 & 0.688 &	88\% ( $^4$I)$^5$K~ +~8\% ( $^4$I)$^3$I +~3\% ($^2_2$H)$^3$I&\\   
    19616.608& - 69 & 0.869 &	53\% ( $^4$I)$^3$I~ +15\% ( $^4$I)$^5$I +~9\% ( $^4$I)$^5$K &\\   
    20348.715&   29 & 0.924 &	76\% ( $^4$I)$^5$I~ +12\% ( $^4$I)$^3$H +~3\% ( $^4$I)$^3$I &\\   
    22181.265& -  8 & 1.032 &	58\% ( $^4$I)$^5$H~ +17\% ( $^4$I)$^3$I +~8\% ( $^4$I)$^3$H &\\   
    23347.692&   61 & 1.153 &	41\% ( $^4$I)$^5$G~ +18\% ( $^4$I)$^3$H +17\% ( $^4$I)$^5$H &\\   
    24878.419& - 12 & 1.141 &	32\% ( $^4$I)$^3$H~ +45\% ( $^4$I)$^5$G +~7\% ( $^4$I)$^5$H &\\   
   (28927)   &      & 1.192 &	80\% ( $^4$I)$^3$G~ +~7\% ( $^4$I)$^5$G +~2\% ( $^4$F)$^3$G &\\  
    29397.378& - 15 & 0.982 &	22\% ( $^4$F)$^5$H~ +19\% ($^2_2$H)$^3$I +~7\% ($^2_1$G)$^3$I    &A\\   
    30175.690& -  7 & 1.051 &	60\% ( $^4$F)$^5$H~ +~7\% ($^2_2$H)$^3$I +~5\% ( $^4$I)$^5$I~~~6s&B \\   
   (31471)   &      & 1.135 &	59\% ( $^4$F)$^5$G~ +13\% ($^2_2$H)$^3$I +~5\% ( $^4$F)$^3$H &\\   
\hline
\multicolumn{5}{l}{$J=6$}	           				 \\
\hline
   (15332)   &      & 0.725 &	93\% ( $^4$I)$^5$L~ +~3\% ($^2_2$H)$^3$K +~3\% ( $^4$I)$^3$K &\\   
    16938.043&   25 & 0.912 &	94\% ( $^4$I)$^5$K~ +~4\% ( $^4$I)$^3$I +~2\% ($^2_2$H)$^3$I &\\   
    20799.314& - 10 & 1.021 &	39\% ( $^4$I)$^3$I~ +22\% ( $^4$I)$^5$I +13\% ( $^4$I)$^3$K  &\\   
    21980.505& - 13 & 1.084 &	70\% ( $^4$I)$^5$I~ +~8\% ( $^4$I)$^3$H +~8\% ( $^4$I)$^3$I  &\\   
    23120.081&   52 & 0.940 &	53\% ( $^4$I)$^3$K~ +14\% ( $^4$I)$^5$H +~9\% ($^2_2$H)$^3$K &\\   
    24058.489& - 50 & 1.122 &	50\% ( $^4$I)$^5$H~ +28\% ( $^4$I)$^3$I +~5\% ( $^4$I)$^3$K  &\\   
    25034.830&   29 & 1.250 &	48\% ( $^4$I)$^5$G~ +22\% ( $^4$I)$^3$H +13\% ( $^4$I)$^5$H  &\\   
   (26724)   &      & 1.234 &	43\% ( $^4$I)$^5$G~ +29\% ( $^4$I)$^3$H +~6\% ( $^4$G)$^3$H  &\\   
   (29509)   &      & 1.002 &	19\% ($^2_2$H)$^3$K~ +15\% ( $^4$I)$^3$K +~9\% ($^2_2$H)$^1$I&\\   
    31146.429& - 30 & 1.060 &	31\% ( $^4$F)$^5$H~ +31\% ($^2_2$H)$^3$K +~8\% ( $^4$F)$^5$G &A\\   
    31559.191&  132 & 1.151 &	43\% ( $^4$F)$^5$H~ +11\% ( $^4$F)$^5$G +10\% ($^2_2$H)$^1$I &B\\   
   (32696)   &      & 1.096 &	24\% ($^2_2$H)$^1$I~ +23\% ( $^4$F)$^5$G +12\% ($^2_2$H)$^3$I&\\   
\hline											     
\newpage										     
\caption{continued.}\\									     
\hline
$E^{\rm a}$ &$\Delta$$E^{\rm b}$ & $g$ &Composition$^{\rm c}$& Remark$^{\rm d}$ \\ 
\hline
\hline
\multicolumn{5}{l}{$J=7$}       					\\
\hline
   (17150)   &      & 0.915 &	96\% ( $^4$I)$^5$L~ +~2\% ($^2_2$H)$^3$K +~1\% ( $^4$I)$^3$K &\\   
    18656.240&   12 & 1.054 &	97\% ( $^4$I)$^5$K~ +~1\% ( $^4$I)$^3$I +~1\% ($^2_2$H)$^3$I &\\   
    22702.749& - 13 & 1.146 &	54\% ( $^4$I)$^5$I~ +22\% ( $^4$I)$^3$I +14\% ( $^4$I)$^3$K  &A\\   
    24003.190& - 12 & 1.176 &	35\% ( $^4$I)$^5$I~ +32\% ( $^4$I)$^3$I +19\% ( $^4$I)$^5$H  &B\\   
   (24602)   &      & 0.891 &	81\% ( $^4$I)$^3$L~ +~8\% ( $^2$K)$^3$L +~3\% ( $^4$I)$^3$K  &\\   
    25319.572&   57 & 1.099 &	49\% ( $^4$I)$^3$K~ +26\% ( $^4$I)$^5$H +~9\% ($^2_2$H)$^3$K &\\   
    26097.872& - 63 & 1.186 &	42\% ( $^4$I)$^5$H~ +31\% ( $^4$I)$^3$I +12\% ( $^4$I)$^3$K  &\\   
    31781.776& - 52 & 1.141 &	33\% ( $^4$F)$^5$H~ +27\% ($^2_2$H)$^3$K +~9\% ($^2_1$G)$^3$I&A\\   
    32832.419& - 39 & 1.155 &	45\% ( $^4$F)$^5$H~ +30\% ($^2_2$H)$^3$K +~6\% ( $^2$K)$^3$K &B\\   
\hline											     
\multicolumn{5}{l}{$J=8$}	 					\\
\hline
   (19082)   &      & 1.043 &	99\% ( $^4$I)$^5$L~ +~1\% ($^2_2$H)$^3$K	   \\
    20410.864& - 06 & 1.150 &	97\% ( $^4$I)$^5$K~ +~1\% ( $^4$I)$^3$L + ~1\% ( $^2$K)$^3$L &\\  
    24592.234& - 67 & 1.229 &	83\% ( $^4$I)$^5$I~ +14\% ( $^4$I)$^3$K + ~1\% ( $^2$K)$^3$K &\\   
   (26769)   &      & 1.019 &	84\% ( $^4$I)$^3$L~ +~8\% ( $^2$K)$^3$L + ~2\% ( $^4$I)$^5$K &\\   
    27441.879&   77 & 1.140 &	67\% ( $^4$I)$^3$K~ +13\% ( $^4$I)$^5$I +12\% ($^2_2$H)$^3$K &\\   
\hline											     
\multicolumn{5}{l}{$J=9$}	 	   	 			 \\
\hline
   (21104)   &      & 1.133 &	99\% ( $^4$I)$^5$L~				  \\
    22197.090& - 28 & 1.218 &	95\% ( $^4$I)$^5$K~ + ~3\% ( $^4$I)$^3$L + ~2\% ( $^2$K)$^3$L& \\ 
   (28963)   &      & 1.114 &	83\% ( $^4$I)$^3$L~ + ~8\% ( $^2$K)$^3$L + ~4\% ( $^4$I)$^5$K& \\ 
\hline											     
\multicolumn{5}{l}{$J=10$}	           				 \\ 		     
\hline											     
   (23200)   &      & 1.200 &   99\% ( $^4$I)$^5$L~ + ~1\% ( $^2$K)$^3$M		 &\\ 
\hline	     	 													
\end{longtable}
	\end{center}												      
$^a$ Calculated values for unknown levels are given in brackets \\
$^b$ Differences between observed and calculated level energies \\
$^c$ 6s -- contribution from the 4f$^3$6s configuration.\\
$^d$ Letters A and B are used to distinguish levels with the same leading term.
	\end{footnotesize}											      

\end{document}